\documentclass[10pt,a4paper]{article}
\usepackage{arxiv}
\usepackage{tikz}
\usepackage[utf8]{inputenc} % allow utf-8 input
\usepackage[T1]{fontenc}    % use 8-bit T1 fonts
\usepackage{hyperref}       % hyperlinks
\usepackage{url}            % simple URL typesetting
\usepackage{booktabs}       % professional-quality tables
\usepackage{amsfonts}       % blackboard math symbols
\usepackage{nicefrac}       % compact symbols for 1/2, etc.
\usepackage{microtype}      % microtypography
\usepackage{lipsum}
\usepackage{miller}

\usepackage{amssymb}
\usepackage{lineno,hyperref}
\usepackage{amsmath}
\usepackage{bm}
\usepackage{subcaption}
\usepackage{graphicx}
\usepackage{svg}
\usepackage{rotating}
\usepackage{lscape}
\usepackage{array,multirow}
\usepackage[title]{appendix}
\usepackage{float}
\usepackage{stmaryrd}
\usepackage{textcomp}
\usepackage{algorithm}
\usepackage{algpseudocode}
\usepackage{amsthm}
\usepackage{listings}
\usepackage{xcolor}
\usepackage{siunitx}
\usepackage{pifont}
\usepackage{lineno}
\usepackage{comment}
\usepackage{import}
\usepackage{physics}
\usepackage{comment}
\usetikzlibrary{positioning}

    \DeclareFontFamily{OT1}{pzc}{}
\DeclareFontShape{OT1}{pzc}{m}{it}{<-> s * [1.10] pzcmi7t}{}
\DeclareMathAlphabet{\mathpzc}{OT1}{pzc}{m}{it}
\usepackage{natbib}
\setcitestyle{authoryear}

\usepackage[algo2e,ruled,vlined]{algorithm2e}
\DontPrintSemicolon

\newtheorem{remark}{Remark}

\DeclareMathOperator*{\argmin}{arg\,min}

 \title{Machine-learning convex and texture-dependent macroscopic yield from crystal plasticity simulations}
\author{
  Jan N. Fuhg \\
Sibley School of Mechanical and Aerospace Engineering\\ Cornell University\\ Ithaca, NY, United States\\
  \And
  Lloyd van Wees \\
  Department of Mechanical Engineering \\
  The University of Alabama \\
  Tuscaloosa, AL, United States \\
  \And
  Mark Obstalecki\\
  Air Force Research Laboratory\\ Materials and Manufacturing Directorate\\
  Wright-Patterson AFB, OH, United States\\
  \And
    Paul Shade\\
  Air Force Research Laboratory\\ Materials and Manufacturing Directorate\\
  Wright-Patterson AFB, OH, United States\\
   \And
 Nikolaos Bouklas \\
  Sibley School of Mechanical and Aerospace Engineering\\
  Center for Applied Mathematics\\
  Cornell University \\
Ithaca, NY, United States \\
   \And
   Matthew Kasemer\\
     Department of Mechanical Engineering \\
  The University of Alabama \\
  Tuscaloosa, AL, United States \\
  }

\begin{document}
\maketitle

\begin{abstract}
The influence of the microstructure of a polycrystalline material on its macroscopic deformation response is still one of the major problems in materials engineering. For materials characterized by elastic-plastic deformation responses, predictive computational models to characterize crystal-plasticity (CP) have been developed. However, due to their large demand of computational resources, CP simulations cannot be straightforwardly implemented in hierarchical computational models such as FE$^{2}$. This bottleneck intensifies the need for the development of macroscopic simulation tools that can be directly informed by microstructural quantities. Using a 3D Finite-Element solver for CP, we generate a macroscopic yield function database based on general loading conditions and crystallographic texture. We furthermore assume an independence of the yield function to hydrostatic pressure of the yield function. Leveraging the advancement in statistical modeling we describe and apply a machine learning framework for predicting macroscopic yield as a function of crystallographic texture. The convexity of the data-driven yield function is guaranteed by using partially input convex neural networks as the predictive tool. Furthermore, in order to allow for the predicted yield function to be directly incorporated in time-integration schemes, as needed for the Finite Element method, the yield surfaces are interpreted as the boundaries of signed distance function level sets.

\end{abstract}

\keywords{Crystal-plasticity  \and Data-Driven Yield Function \and Multiscale \and Polycrystals \and Physics-informed constraints}

\section{Introduction}\label{sec::1}
The study of polycrystals such as metals, ceramics, polymers and rocks has long been of interest for a variety of applications in engineering and materials science. 
Understanding their response, however, is complicated  
due to the fact that
most major quantities of interest, such as strength or electrical conductivity of polycrystals, are anisotropic.
The degree of anisotropy is majorly influenced by the preferred orientation of crystallites (or texture) \citep{d1838introduction}. 
For example in the extreme case of complete randomness of orientations, the macroscopic properties will be isotropic. On the other hand, a representative volume element characterized by a single preferred orientations will inherit the anisotropic behavior of the perfect single crystal \citep{wenk2004texture} .

The texture of polycrystals also effects their plastic behavior in particular the current state and the evolution of plastic flow, i.e. sharp textures lead to predictions of sharp vertices \citep{kocks1998texture}.
Thanks to the early works of \cite{taylor1938plastic} and the later works of \cite{schmid1950plasticity} it is generally known that the gliding of dislocations due to shearing on selected crystallographic planes is the reason
for plastic yielding of single crystals. Even though these processes are relatively well understood,  the prediction of texture-dependent plastic behavior for polycrystals is still an area of active research.
In particular, fitting and predicting macroscopic yield surfaces in polycrystalline materials is of extensive interest, see \cite{plunkett2006anisotropic, watanabe2010method}. %\\
This process is made more complicated by (usually) requiring the yield function to be convex, which goes back to the ideas established in \cite{drucker1951more,drucker1959definition}.

When enough data of the macroscopic yield surface is available, mathematical models can be designed to predict the onset of plastic flow.
Over the years a lot of effort has focused into developing  phenomenological yield and failure criteria for materials.
Phenomenological yield functions do not result directly from microstructure-based calculations \citep{barlat1991six} but use user-defined combinations of simple mathematical expressions (linear, polynomial, log,...) to describe the observed phenomena. 
For metals, where the studies of \cite{bridgman1923compressibility,Bridgman1952} paved the way for the general assumption that the yield function is independent of the hydrostatic pressure, different phenomenological models have been developed. Going back to the simple but powerful formulations of \cite{tresca1864ecoulement} and \cite{mises1913mechanik} the complexity has gradually increased, see e.g. \cite{hencky1924theorie,burzynski1929anstrengungshypothesen,hershey1954plasticity,hosford1972generalized,hecker1976experimental}. However, these early works assumed generally isotropic behavior. 

On the other hand, \cite{hill1948theory} and \cite{hoffman1967brittle} proposed fundamental anisotropic phenomenological yield surfaces.
More recently, Barlat and his coworkers studied and proposed anisotropic yield functions for two and three-dimensional application \cite{barlat1997yielding,barlat2003plane,barlat2005linear,aretz2013new}. The yield surfaces employed in Barlat's works are variations of the function used in \cite{hosford1972generalized} which is convex by design but isotropic.
However, these models make use of the isotropic plasticity equivalent stress transformation introduced by 
\cite{karafillis1993general} which employs a fourth order tensor to potentially allow the representation of any anisotropic state of a material. This is achieved by defining convexity-preserving linear operator which maps the fourth order tensor onto the actual stress tensor. Thereby, the components of the stress tensor are "weighted" according to the anisotropy present in the material. 
%\cite{soare2016modeling} represent symmetric and asymmetric yield functions by Fourier series while preserving their convexity.  

Even though these proposed and commonly applied phenomenological models show proficient results for a variety of different applications \cite{grytten2008evaluation,plunkett2008orthotropic,banabic2010advances,esmaeilpour2018calibration}
they have some severe drawbacks which include but are not limited to the following:
\begin{itemize}
    \item They rely extensively on user experience, intuition, knowledge of material symmetries and tensor-value function theory. 
    \item The parameters used to describe anisotropic behavior, introduced by isotropic plasticity equivalent stress transformation, are rarely physically motivated, i.e. texture characteristics such as its character and spread are not utilized directly.
    \item Conditions for the convexity of a phenomenological model need to be established and their fulfillment significantly limits the design space of these models.
    \item The functional dependency between the yield function and its arguments (loading condition, texture, ...) has to be simple enough to allow for a proficient fit of the data by only combining a limited number of user-chosen mathematical expressions.
    \item When the complexity of the yield function increases, the number of unknown parameters that need to be fit also increase, hindering the fitting process.
\end{itemize}
However, the availability of closed form models for anisotropic yield surfaces is crucial in using the finite element analysis on a structural level \citep{de2011computational}. 
Hence, instead of using phenomenological models we resort to methods that are able to fit the available data in an automated fashion using techniques from computational statistics but still allow for explicitly and analytically available yield surfaces.

In this context data-driven or machine learning (ML) approaches have been an emergent tool in the computational sciences in recent years.
In this context, machine learning has been used to directly solve forward and inverse problems involving partial differential equations (\cite{raissi2019physics,FUHG2021110839,fuhg2021interval}). ML has also been utilized for the development of intrusive and non-intrusive Reduced Order Modeling (ROM) schemes for accelerated solutions of PDEs \citep{kadeethum2021non,hernandez2021deep, kadeethum2021framework}

Recently, data-driven models have been employed to bypass the use of phenomenological constitutive models by introducing data-driven constitutive modeling (recapitulating strain-stress relationships) in solid mechanics
\citep{huang2020machine, fuhg2021modeldatadriven, fuhg2021local,fuhg2021physics} as a means to enable hierarchical multiscale calculations as well as direct use of experimental data. 
\cite{vlassis2021sobolev,vlassis2021component} proposed a component-based data-driven model for elastoplastic materials where the yield function is trained separately from the elastic response by using a level-set approach.

Based on these ideas we study the development of data-driven  texture-dependent yield surfaces that maintain convexity; texture-based parameters are directly used as an input to a machine learning model.
Leveraging datasets from CP finite element simulation to obtain averaged macroscopic yield functions, we rely on level-set methods to obtain a predictive tool that can directly be utilized in a time-integration loop in structural finite element problems. In order to ensure the convexity of the trained yield surface we employ partially input convex neural networks (pICNN) proposed by \cite{amos2017input} to train our datasets. In the context of hyperelastic material modeling in solid mechanics a simplification of these networks has been used in \cite{klein2022polyconvex}.

The organization of the paper is as follows. We first provide a detailed account of CP finite element modeling in Section \ref{sec::2}. The frameworks for sample design and the simulations are shortly introduced in Section \ref{sec::3}. In Section \ref{sec::4} the steps for preprocessing the data are described. 
Neural networks and the pICNN formulation are summarized in Section \ref{sec::4}.
The main results of the paper including a study for interpolation and extrapolation capabilities are discussed in Section \ref{sec::5}.
The paper is concluded in \ref{sec::6}.

\section{Crystal-plasticity finite element modeling}\label{sec::2}

Crystal plasticity finite element modeling (CPFEM) has emerged in the last two decades as an adept tool at predicting both micro- and macroscopic behaviors of polycrystalline alloys~(\cite{marin_1998_1, marin_1998_2,roters}). In the last decade, particularly, the modeling community has benefited from an increase in both computational power and technical ability, such that it is now commonplace to perform simulations on high-fidelity representations of microstructures, such that the effect of grain morphology, neighborhoods, and texture on the development of plasticity both locally and globally may be inspected~(\cite{kasemer_beta_2017, cappola}). Various studies have employed these high-fidelity capabilities to help determine material parameters~(\cite{Wielewski2017, Dawson2018}), lattice reorientation~(\cite{Quey2012, Quey2015}), texture evolution~(\cite{kasemer_wenk}), and generally the development of plasticity~(\cite{kasemer_slip_2017}).

Modern CPFEM simulations generally consider explicit three-dimensional representations of microstructures---i.e., the grain morphology, intra-grain orientations (and thus the crystallographic texture), and generalized loading conditions (allowing for testing of, generally, triaxial-principal loading conditions). Broadly, CPFEM simulations are able to predict both intra-grain behavior, as well as the behavior of the entire domain of interest (i.e., macroscopic behavior). Of particular note, the development of plasticity may be tracked. This allows for an understanding both of how plasticity evolves at the level of individual grains, as well as the understanding of macroscopic yield. Coupled with generalized loading conditions, multiple simulations may be performed in an effort to elucidate an envelope of the macroscopic yield surface (as described in detail in Section~\ref{subsec::simsuite}).

The crystal plasticity finite element solver employed in this study, FEPX~(\cite{fepx_arxiv}), employs anisotropic elasticity and plasticity. The models that FEPX considers are ductile and isothermal (i.e., we do not consider fracture or thermal strain models). The models are embedded in a non-linear finite element solver. For sake of brevity, description of the finite element implementation is omitted, the details of which can be found elsewhere~(\cite{fepx_arxiv, marin_1998_1, marin_1998_2}), and the description of the kinematics, models, and evolution equations is truncated to highlight the primary points of interest.

The total deformation response of an element in a finite element mesh can be described using the deformation gradient. In this formulation, the deformation gradient is split into an elastic portion, a rotation, and a plastic portion:
    \begin{equation}
        {\bf F} = {\bf F}^e {\bf F}^R {\bf F}^p
    \end{equation}

Elasticity is considered via Hooke's law:
    \begin{equation}
        \bm{\sigma} = \mathcal{C}\left({\bf r}\right) \bm{\epsilon}
    \end{equation}
where the stress, $\bm{\sigma}$, is related linearly to the strain, $\bm{\epsilon}$, via the anisotropic elastic stiffness tensor, $\mathcal{C}$, which is reduced to reflect major, minor, and crystal symmetry~(\cite{bower, nye, hosford}). The stiffness tensor is a function of the orientation of the crystal, ${\bf r}$, parameterized as a Rodrigues vector~(\cite{frank, kumar}).

Considering plasticity, the slip system rate of shear is governed by a rate-dependent phenomenological power-law model:
    \begin{equation}
        \dot{\gamma}^{k} = \dot{\gamma}_0 \left( \frac{\left| \tau^k \right|}{\tau_c} \right)^{\frac{1}{m}} \hbox{sgn}\left( \tau^{k} \right)
    \end{equation}
where the slip rate $\dot{\gamma}$ on the $k$-th slip system is related primarily to the resolved shear stress, $\tau$, on that slip system, and the critical resolved shear stress, $\tau_c$, where rate-dependence is controlled via the power parameter, $m$. The resolved shear stress is calculated as:
    \begin{equation}
        \tau^k = \bm{\sigma} : {\bf P}^k
    \end{equation}
where ${\bf P}$ is the symmetric portion of the Schmid tensor~(\cite{schmid1935kristallplastizitat}), calculated as the dyadic product between the slip direction and the slip plane normal.

The critical resolved shear stress is permitted to evolve as a a function of the amount of accumulated plastic shear:
    \begin{equation}
        \dot{\tau}^k_c = h_0 \left( \frac{\tau_s - \tau^k}{\tau_s - \tau_c} \right) \dot{\Gamma}
    \end{equation}
where $\tau_s$ is the saturation value for the critical resolved shear stress, and $\dot{\Gamma}$ is the sum of the shears on all slip systems for a given element:
    \begin{equation}
        \dot{\Gamma} = \sum^k{\dot{\gamma}^k}
    \end{equation}
    
The crystal experiences a rotation, or reorientation:
    \begin{equation}
        \dot{\bf r} = \frac{1}{2} \left( \bm{\omega} + \left(\bm{\omega} \cdot {\bf r} \right) {\bf r} + \bm{\omega} \times {\bf r} \right)
    \end{equation}
where $\bm{\omega}$ is the lattice spin, is based on the plastic spin rate tensor~(\cite{marin_1998_1}).

\section{CPFEM Sample Generation and Simulation Suite}\label{sec::SampleGeneration}

\subsection{Microstructure generation}

To generate virtual representations of microstructures and concomitant meshes for use in CPFEM simulations, we use the software package Neper~(\cite{neper}). Neper is capable of generating large microstructural representations with defined grain size and shape distributions, via a Laguerre tessellation~(\cite{kasemer_beta_2017}). This is broadly controlled through optimization of distributions of the grain size and shape to user-defined target distributions. The grain size is defined via the metric of the normalized equivalent grain diameter, or the diameter of the sphere of equal volume to the grain, normalized by the average of diameters of all grains. The shape is defined via the metric of sphericity, defined as the ratio of the surface area of the grain to the surface area of the sphere of equivalent volume to the grain.

\begin{figure}[h!]
\begin{subfigure}[b]{.5\linewidth}
\centering
    \centering
    \includegraphics[scale=0.2]{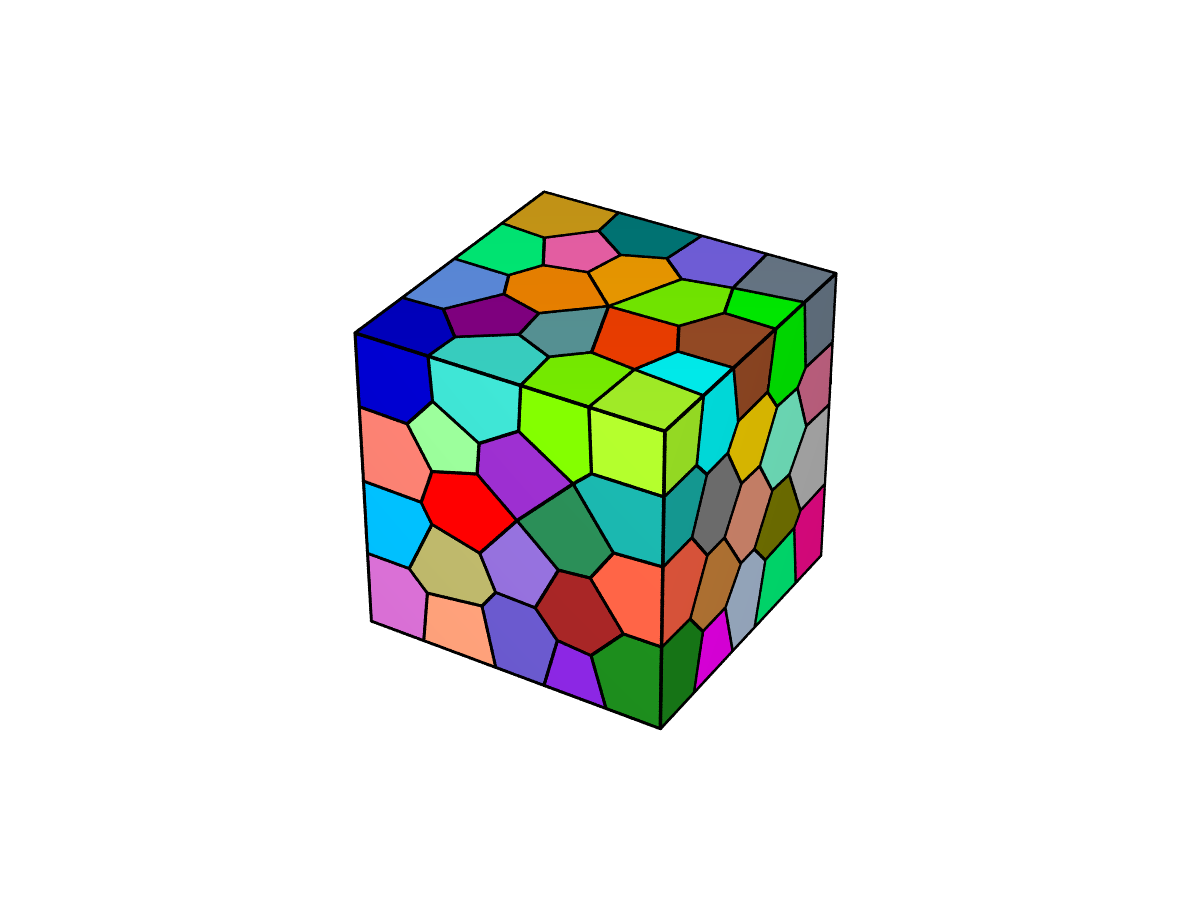}
    \caption{Tessellation}\label{fig::100grains_tess}
\end{subfigure}%
\begin{subfigure}[b]{.5\linewidth}
\centering
    \centering
    \includegraphics[scale=0.2]{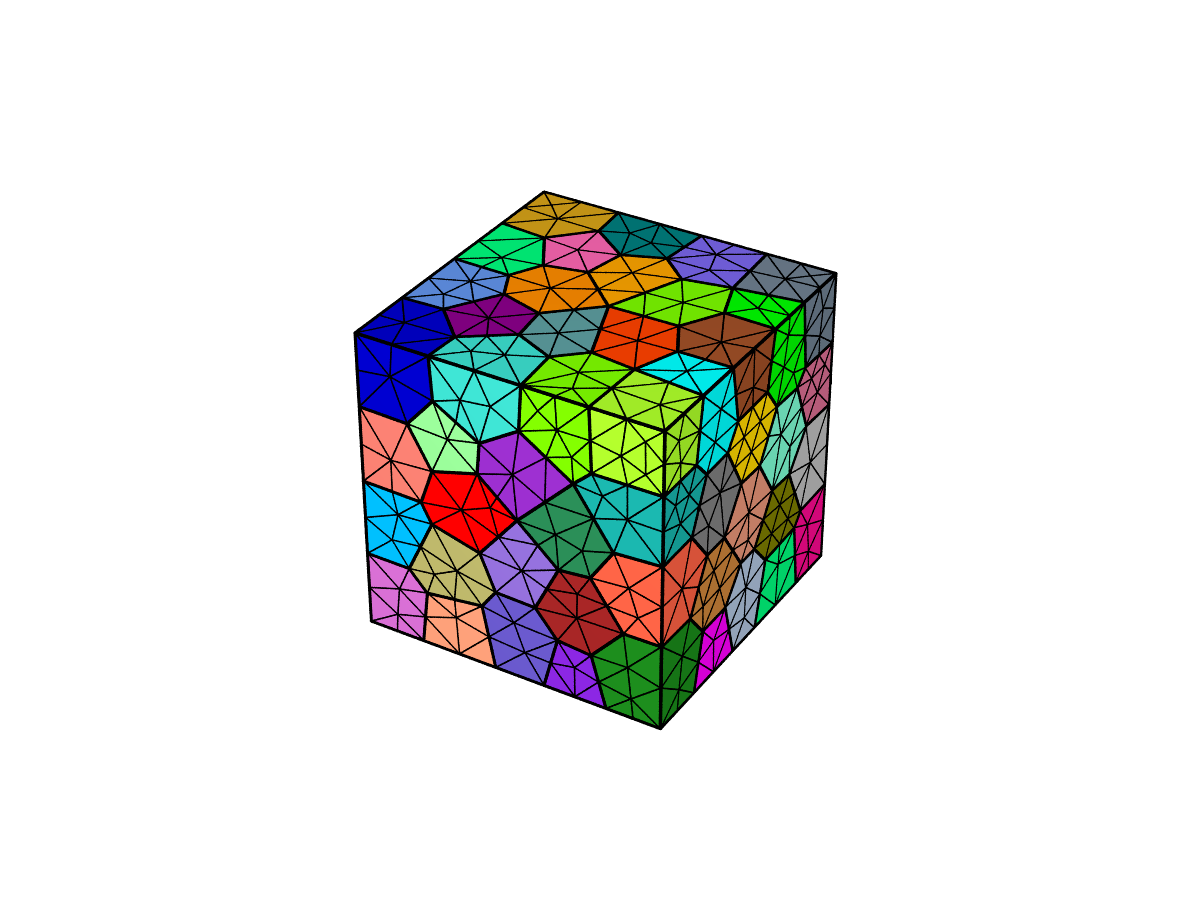}
    \caption{Mesh}\label{fig::100grains_mesh}
\end{subfigure}%
    \caption{100 grain polycrystalline domain with concomitant finite element mesh. Each grain is colored arbitrarily.}
    \label{fig:100grains}
\end{figure}

With focus on the effect of texture on the macroscopic yield surface, we aim to minimize the influence of the geometric morphology of the microstructure on the macroscopic yield response of the material. To that end, we choose distributions of grain sizes and shapes to create a highly equiaxed microstructure---i.e., a microstructure with grains of relatively constant size and shape. To achieve this, we use a Dirac distributions for both the normalized equivalent diameter and the sphericity. A polycrystalline sample with 100 grains is generated, with a mesh comprised of approximately 10,000 elements (i.e., 100 elements per grain). The number of grains is chosen to limit the computational expense, which facilitates the rapid generation of large datasets necessary for training. Qualitatively, this produces a fairly geometrically homogeneous microstructure, plotted in Figure~\ref{fig:100grains}. The geometric morphology of the microstructure and the finite element mesh are held fixed across all simulations.

\subsection{Texture generation and parameterization} \label{sec:::texture}

Crystallographic texture is considered by applying crystallographic orientations to grains (specifically the elements within a grain) such that the overall distribution of orientations adheres to a desired texture. In this study, grains are initially single crystals (i.e., no orientation spread), such that all elements that belong to a grain are assigned the same initial orientation. Neper is again employed to generate orientations for a user-defined texture. Neper considers the definition of texture via the texture character (i.e., the average orientation for a distribution), the magnitude of orientation spread for a specific texture character, and an assumed peak shape. Generally, this can be a combination of texture peaks or fibers, the total distribution being a sum (normalized) of the individually defined distributions. Ultimately, the texture (and for the sake of this study, the polycrystal) is fully described as the relative weight of each texture component (of known character; i.e., cube, Goss, Brass, etc.), and the amount of spread for each individual component:
\begin{equation}\label{eq::Texture}
    \bm{\mathcal{T}} = \{ \mathcal{T}^w_1, \mathcal{T}^s_1, \mathcal{T}^w_2, \mathcal{T}^s_2, ..., \mathcal{T}^w_t, \mathcal{T}^s_t \}
\end{equation}
where $\mathcal{T}^w_b$ signifies the weight, and $\mathcal{T}^s_n$ the spread, of the $t$-th texture component. While more sophisticated and generalized methods exist to parameterize ODFs such as the employ of discrete spherical harmonics~(\cite{Wielewski2017,Dawson2018}), the method described here requires (generally) lower dimensionality, may be expanded to consider as few or as many texture components as necessary, and is able to sufficiently describe ODFs that are known to exist for common processing routes.

Here, we focus on a singular texture character rather than the general case of combinations of various texture components. Specifically, we focus on an orientation distribution where the average orientation is centered at the cube texture component~(\cite{Raabe2004}) (i.e., a crystal with no rotation from the sample coordinate system, or ${\bf r} = \left( 0,0,0 \right)$). The distribution of orientations around this texture component is assumed to be a normal distribution, and the amount of spread is defined as $\theta_m$, or the average amount of misorientation from the nominal peak average. Nine different textures are considered, ranging from $\theta_m = 5^\circ$ to $\theta_m = 25^\circ$ in steps of $2.5^\circ$. Thus, the material description always follows the form $\bm{\mathcal{T}} = \{ \theta_m \}$ (since the weight of this texture component is necessarily 1, and is functionally arbitrary when considering only a single texture component).

\begin{figure}[ht!]
\begin{subfigure}[b]{0.33\linewidth}
\centering
    \centering
    \includegraphics[height=2.5in]{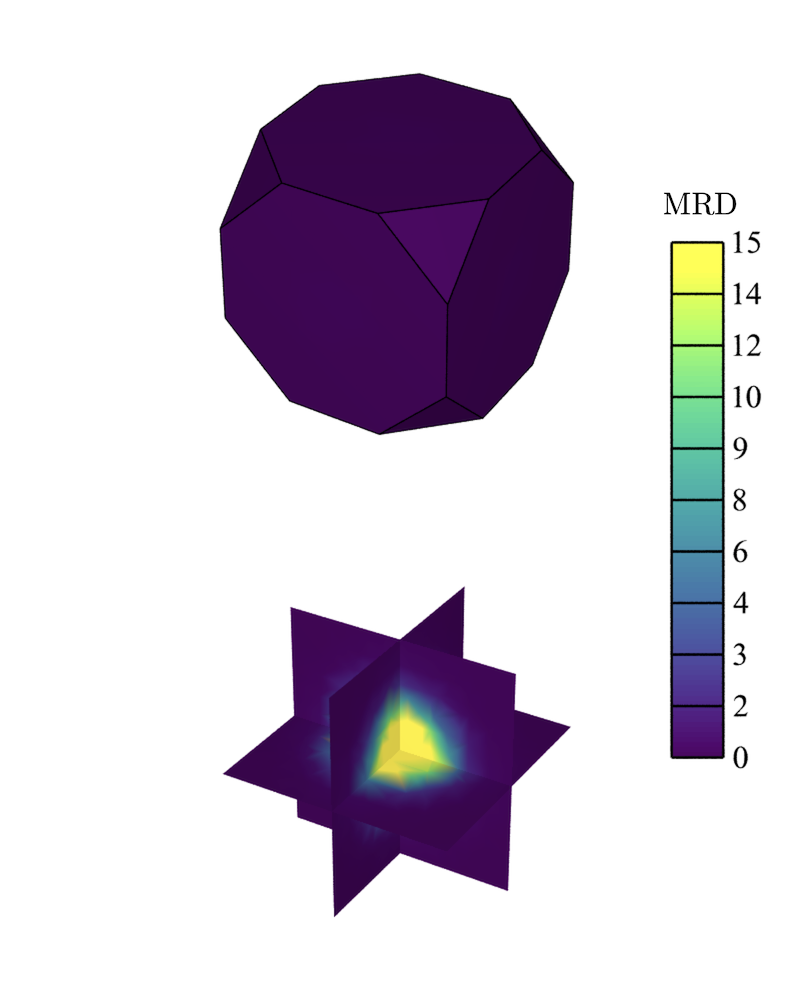}
    \caption{$\theta_m=5^\circ$}\label{fig::odf_5}
\end{subfigure}%
\begin{subfigure}[b]{0.33\linewidth}
\centering
    \centering
    \includegraphics[height=2.5in]{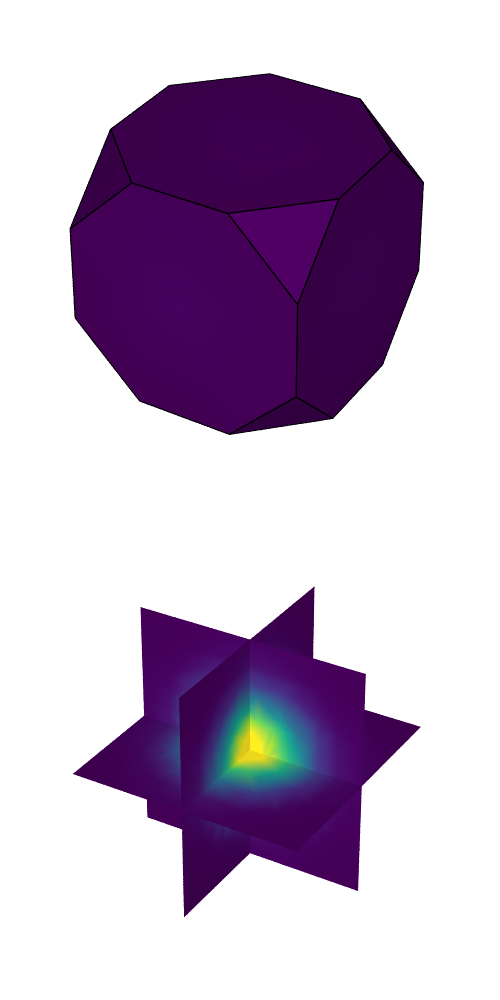}
    \caption{$\theta_m=15^\circ$}\label{fig::odf_15}
\end{subfigure}%
\begin{subfigure}[b]{0.33\linewidth}
\centering
    \centering
    \includegraphics[height=2.5in]{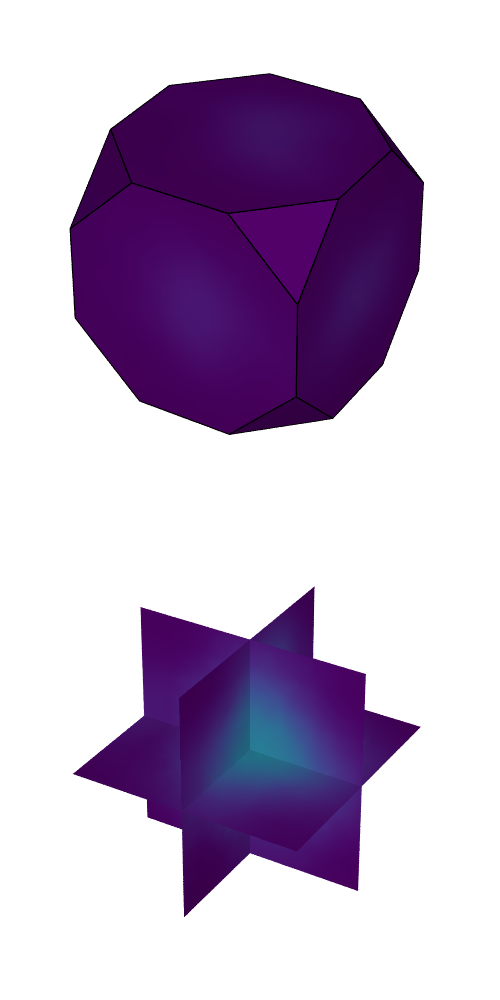}
    \caption{$\theta_m=25^\circ$}\label{fig::odf_25}
\end{subfigure}%
    \caption{Orientation distribution functions for a single texture component (cube component), for average misorientations of (\protect{\subref{fig::odf_5}}) $\theta_m=5^\circ$, (\protect{\subref{fig::odf_15}}) $\theta_m=15^\circ$, and (\protect{\subref{fig::odf_25}}) $\theta_m=25^\circ$. ODFs are plotted in the cubic fundamental region of Rodrigues orientation space, and each ODF is shown on the surface of the fundamental region (top images), as well as slices of isosurfaces of the interior of the fundamental region (bottom images). Scale in (\protect{\subref{fig::odf_5}}) depicts multiples of random distribution (MRD), and is constant for all ODFs. Note that the maximum values for each ODF are $33.99$, $17.68$, and $7.43$ for $\theta_m=5^\circ$, $15^\circ$, and $25^\circ$, respectively, and thus the plots may clip (or saturate) if their maximum value is over the scale bar, whose range is chosen to demonstrate the change in character and intensity.}
    \label{fig:odfs}
\end{figure}

Figure~\ref{fig:odfs} shows three representative orientation distribution functions (ODF) used to generate orientations for use in simulations, plotted in the cubic fundamental region of Rodrigues space. Qualitatively, we note that each ODF is a normal distribution with its peak centered at the cube texture component. As $\theta_m$ increases, the ODF becomes more diffuse over the fundamental region---that is, the orientations become more random, or less preferentially oriented to the cube texture component, though there still exists some non-random texture.

It is worth noting that---while this study focuses solely on the effect that crystallographic texture has on the macroscopic yield surface (and thus the material description, $\bm{\mathcal{T}}$, contains information only about the texture)---other descriptors may also be included to describe the state of the material: ranging from geometric description of the microstructure (grain size / shape distributions) to elastic/plastic modeling parameters. In other words, the framework described here can allow for increasingly-complex material descriptions.

\subsection{Material selection}

To contextualize the choice in material, we must first discuss the implications of the model, model parameters, and the resulting macroscopic yield surface. 

Since the amount of plastic evolution (specifically hardening) at the point of macroscopic yield is expected to be minimal, the choice of plasticity parameters (within reasonable bounds) is largely arbitrary, as they will not greatly influence the shape of the (initial) macroscopic yield surface. Likewise, since the shape of the yield surface will be primarily a function of the geometry of the slip modes in the polycrystal (and the relative strength between slip families, should multiple families be necessary or considered), the initial slip system strength will not change the shape of the yield surface, but only the magnitude at which the material will yield. Consequently, for a given crystal symmetry, the critical resolved shear stress is again largely arbitrary, and determines only the magnitude at which yield occurs (i.e., all other aspects fixed, the yield surface will be the same shape for two samples with differing initial critical resolved shear stresses, simply dilated for the sample with the higher value). Note that this is true for our pursuit of the initial yield surface. Understanding the evolution of the yield surface would require careful selection of modeling parameters and training for each new set of parameters, dependent on the degree of sensitivity of macroscopic behavior to the changes in parameters.

For the study at hand, we opt for a material with a cubic crystal structure (face-centered cubic, or FCC) to reduce complexity due to the potential necessity of multiple slip families of disparate strengths (e.g., as would be expected for materials with hexagonal/HCP crystal structure). We assume room temperature behavior, such that only the \hkl{1 1 1} \hkl<1 1 0> slip family is expected to be active. Additionally, a material with an appreciable amount of single crystal anisotropy will yield more pronounced differences in the behavior of the macroscopic yield surfaces, and is thus prioritized. With these collective points considered, we choose Okegawa mold copper (OMC) copper~(\cite{wong}) as our model material system. Possessing an FCC crystal type which exhibits slip on a single family (at room temperature), the single crystal of OMC copper exhibits an appreciable amount of elastic anisotropy, with a Zener ratio of approximately 3.2~(\cite{bower, wong}). Further, while the plasticity parameters are largely arbitrary, there exists a high degree of confidence in the elastic and plastic modeling parameters for this material system~(\cite{wong, obstalecki}). These parameters are summarized in Tables~\ref{tab:elastic}~and~\ref{tab:plastic}.

\begin{table}[htbp!]
    \centering
    \begin{tabular}{c c c}
    
    \hline
    {\bf $C_{11}$ (\SI{}{\giga\pascal})} & {\bf $C_{12}$ (\SI{}{\giga\pascal})} & {\bf $C_{44}$ (\SI{}{\giga\pascal})} \\
    \hline
    164 & 122 & 75 \\
    \hline

    \end{tabular}
    \caption{Single crystal elastic constants (as determined in~\cite{wong}).}
    \label{tab:elastic}
\end{table}

\begin{table}[htbp!]
    \centering
    \begin{tabular}{c c c c c}
    
    \hline
    {\bf $\dot{\gamma}_0$ (-) } & {\bf $m$ (-)} & {\bf $h_0$ (\SI{}{\mega\pascal})} & {\bf $\tau_{0}$ (\SI{}{\mega\pascal})} & {\bf $\tau_s$ (\SI{}{\mega\pascal})} \\
    \hline
    1 & 0.01 & 800 & 85 & 285 \\
    \hline

    \end{tabular}
    \caption{Plastic modeling parameters (as determined in~\cite{wong}).}
    \label{tab:plastic}
\end{table}

\subsection{Simulation Suite and dataset} \label{subsec::simsuite}

In general we expect the macroscopic yield function $f$ to be a possibly non-smooth tensor function of the macroscopic applied Cauchy stress components  $\bm{\sigma}_{M}$ and some texture defining parameters $\bm{\mathcal{T}}$ as defined in eq. (\ref{eq::Texture}).

We can then characterize the elastic domain $\mathcal{E}$
by a negative yield function value
\begin{equation}
    \mathcal{E} = \lbrace \bm{\sigma}_{M} \times \bm{\mathcal{T}} \in \mathbb{R}^{3 \times 3} \times \mathbb{R}^{t}| f(\bm{\sigma}_{M}, \bm{\mathcal{T}}) < 0 \rbrace
\end{equation}
with $t\geq 1$.
The boundary $\partial \mathcal{E}$ of $\mathcal{E}$, when the yield function returns zero, signals yielding
\begin{equation}
       \partial \mathcal{E} = \lbrace \bm{\sigma}_{M} \times \bm{\mathcal{T}} \in \mathbb{R}^{3 \times 3} \times \mathbb{R}^{t}| f(\bm{\sigma}_{M}, \bm{\mathcal{T}}) = 0\rbrace.
\end{equation}
Our goal is to approximate the nature of the function $f$ from data. For this, assume for now that for a particular loading condition and a given texture a CPFEM simulation can be run which outputs a binary value indicating if the specimen has yielded or not yielded. 
Hence, by running $N$ simulations in this setting we obtain the data set $\mathcal{D}_{class} = \left\lbrace \left( (\bm{\sigma}^{i}_{M}, \bm{\mathcal{T}}^{i}),\varphi^{i}  \right) \right\rbrace_{i=1}^{N}$ where 
\begin{equation}
    \varphi^{i} = \begin{cases}
    -1, & \text{if yield has occurred} \\
    1, & \text{if elastic}.
    \end{cases}
\end{equation}

While generally triaxial-principal loading is possible (i.e., three orthogonal normal stresses), we focus in this study on biaxial-principal loading conditions (herein referred to simply as `biaxial' loading). Assuming classical behavior where hydrostatic loading does not contribute to yield, and the yield surface remains open in the direction of the hydrostatic axis in principal-stress space ($\sigma_{M, xx} = \sigma_{M, yy} = \sigma_{M, zz}$, assuming no applied shear), the full three-dimensional principal yield surface can be deduced from the biaxial yield surface (since the biaxial yield surface is simply a slice of the three-dimensional yield surface). 
Therefore we assume $f(\sigma_{M,xx}, \sigma_{M, yy}, \bm{\mathcal{T}})$ and can rewrite the established dataset into
\begin{equation}\label{eq::Datasetbinary}
\mathcal{D}_{class, princ} = \left\lbrace \left( ({\sigma}_{M, xx}^{i}, {\sigma}_{M, yy}^{i}, \bm{\mathcal{T}}^{i}),\varphi^{i}  \right) \right\rbrace_{i=1}^{N}.    
\end{equation}

\begin{figure}[h!]
    \centering
    \includegraphics[scale=0.35]{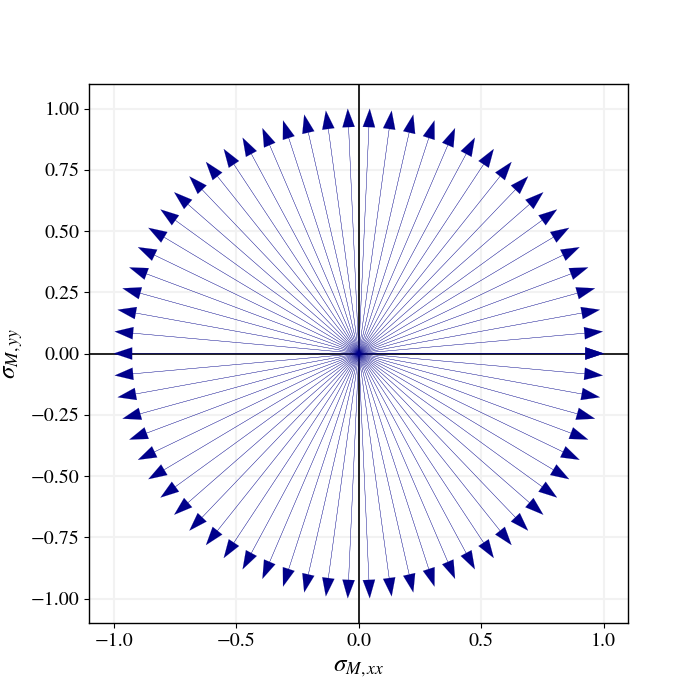}
    \caption{72 loading vectors depicting the ratio of applied loads used in simulations. For each simulation, loads are incrementally increased along one of these vectors (i.e., the ratio of the applied loads $\sigma_{M, xx}$ and $\sigma_{M, yy}$ is held fixed).}
    \label{fig:loadvecs}
\end{figure}

These $N$ simulations need to be able to result in data that is expressive enough to allow for a proficient characterization of the yield function.
In order to achieve this we fill the biaxial plane with a sufficient number of data points to effectively deduce the shape of the yield surface. In this paper, 72 monotonic simulations are performed for a single texture realization with various ratios of applied loads (in this case, $\sigma_{M,xx}$ and $\sigma_{M, yy}$, though for the cube texture considered, the surfaces on which the loads are applied is arbitrary). These load vectors are equally spaced (radially) in the biaxial plane of stress space (i.e., the angle between all applied load vectors is $\frac{\pi}{36}$ radians). The 72 loading vectors applied to the polycrystal are shown in Figure~\ref{fig:loadvecs}.

For each load vector, loads are applied to the surface of the polycrystal such that the ratio of $\sigma_{M, xx}$ to $\sigma_{M, yy}$ is maintained throughout the simulation (within user-defined tolerances). Loads steps are discretized such that they are sufficiently close to one another to resolve the yield point to within 1\% of the macroscopic yield.

As previously stated in Section~\ref{sec:::texture}, 9 different strengths of cube texture are considered. In addition, we perform all simulations on 5 different samples for each texture (i.e., 5 different samples that contain orientations which nominally represent the desired texture). Thus, in total $72 \cdot 9 \cdot 5 = 3,240$ simulations are performed.

To calculate yield, the equivalent (von Mises) stress and equivalent strain behavior is considered. The deviation from linearity is calculated using a $0.1\%$ offset method, and the intersection between the stress-strain curve and the linear offset is determined to be the yield point. The applied loads (stresses) that result in the equivalent stress at yield are calculated. Since data is output at the end of discrete load steps, we employ linear interpolation to find the expected points of yield (discrete load steps are sufficiently close to allow for linearization between load steps), which further allows for binary categorization of discrete load steps as either elastic or plastic. While more sophisticated methods to calculate the onset of plastic yielding exists, \cite{Poshadel2019} demonstrate that while different methods may produce slight differences in their prediction of the onset of macroscopic yield, these differences are minor, and are negligible when it comes to the shape of the yield surface. The invariant von Mises stress is thus appropriate, and readily calculated.

\begin{figure}[ht!]
\begin{subfigure}[b]{.33\linewidth}
\centering
    \centering
    \includegraphics[scale=0.3]{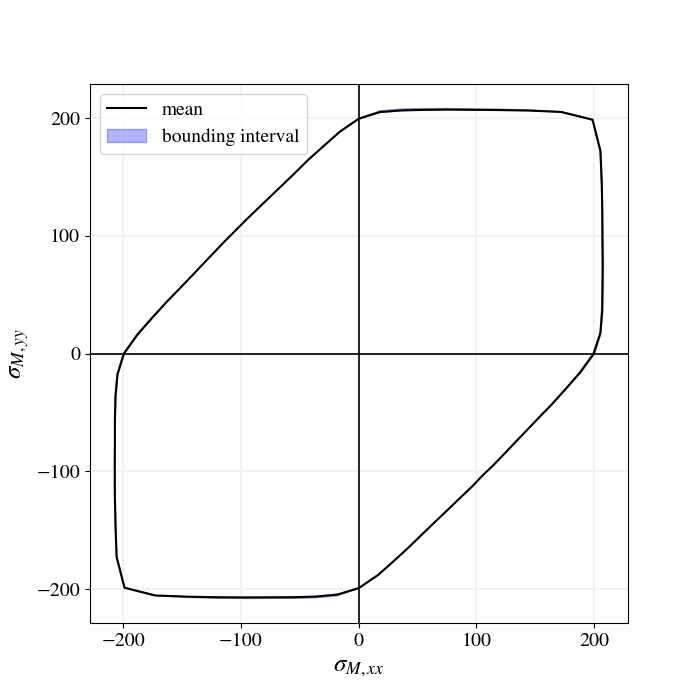}
    \caption{$\theta_{m}=5^{\circ}$}\label{fig::}
\end{subfigure}%
\begin{subfigure}[b]{.33\linewidth}
\centering
    \centering
    \includegraphics[scale=0.3]{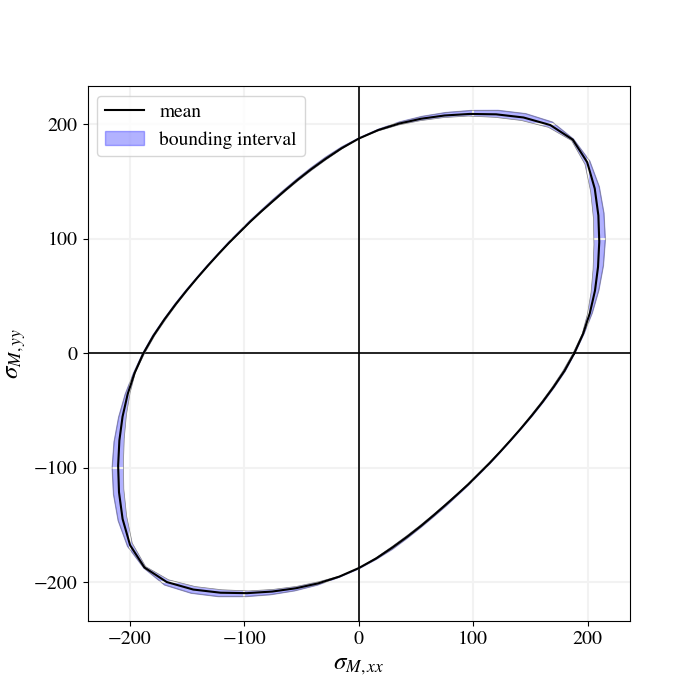}
    \caption{$\theta_{m}=15^{\circ}$}\label{fig::}
\end{subfigure}%
\begin{subfigure}[b]{.33\linewidth}
\centering
    \centering
    \includegraphics[scale=0.3]{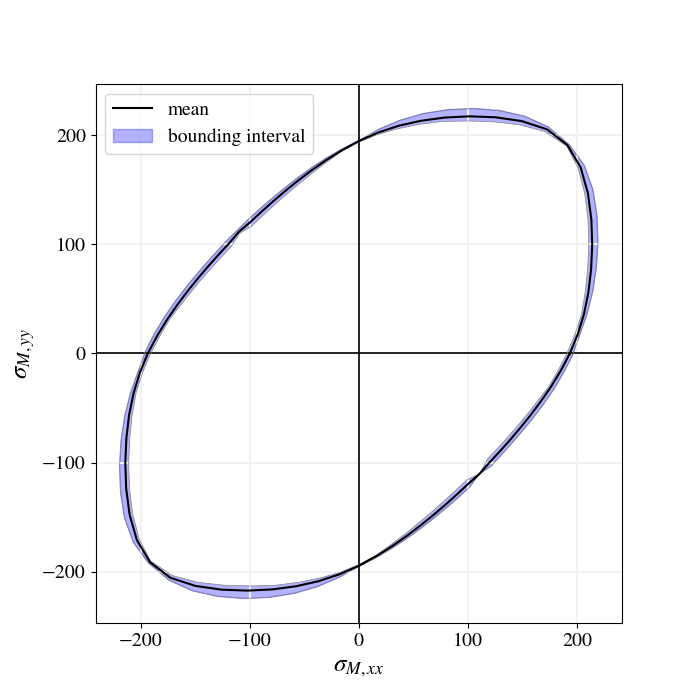}
    \caption{$\theta_{m}=25^{\circ}$}\label{fig::}
\end{subfigure}%
    \caption{Mean and bounding macroscopic yield surfaces of three different spreads of the orientation distribution.}
    \label{fig:RealDataMeanAndBound}
\end{figure}

Yield surfaces are visualized by plotting the applied loads at yield on the biaxial plane in stress-space. Figure~\ref{fig:RealDataMeanAndBound} shows the mean yield and bounding intervals for the yield surfaces generated from the 5 distinct samples for three different textures. Of particular note is how the yield surface is severely faceted for the highly textured sample, while it is much more ellipsoidal for the more randomly textured material (as expected).
In the following only these averaged yield surfaces are used for the subsequent studies of this paper. Thus, we have access to $N=72 \cdot 9= 648$ samples that characterize the yield surfaces. Using a dataset of these points alone is not convenient for training and applying a machine learning based yield function. Hence, in the next section we introduce preprocessing steps that allow us to efficiently train a predictive tool.

\section{Framework for texture-dependent data-driven modeling of yield function}\label{sec::3}

Since the output of the dataset described in eq. \ref{eq::Datasetbinary} is of binary nature (either yield or no yield) it is not smooth. Training the yield surface this way would be problematic because (1) it would not lend itself to be directly used in time-integration schemes like Newton-Raphson \citep{wriggers2008nonlinear} where derivatives of the predicted yield criterion value with regards to the stresses are required, (2) the binary nature of the data leads to under- and overshoot oscillations when using regression-based techniques such as neural networks (see e.g. \cite{zhang2020using} for an example).
In order to circumvent this potential issue, we can instead preprocess the dataset by interpreting the yield boundary based on the framework of level-set theory and actively generate a smooth dataset that allows for seamless training.

Following the approach suggested by \cite{vlassis2021sobolev}, the binary classification dataset $\mathcal{D}_{class, princ}$ is reformulated into a regression problem for each value of $\bm{\mathcal{T}}$ by using the yield boundary $\partial \mathcal{E}$ as the zero level set for the Eikonal equation (see Remark \ref{remark::Nurbs})
  \begin{subequations}
    \begin{align}
    \label{eq::Eikonal}
    \abs{\nabla f(\bm{\mathcal{T}})} &= 1, \qquad \text{for } [\sigma_{M, xx}, \sigma_{M, yy}] \in \mathbb{R}^{2},\\
    \text{subject to }  f &= 0, \qquad \text{for } [\sigma_{M, xx}, \sigma_{M, yy}] \in \partial \mathcal{E}(\bm{\mathcal{T}}).
    \end{align}
  \end{subequations}
The solution of this partial differential equation defines the signed distance function $\phi$ given by
    \begin{equation}
        \phi(\sigma_{M, xx}, \sigma_{M, yy},\bm{\mathcal{T}}) = \begin{cases}
        d(\sigma_{M, xx}, \sigma_{M, yy}), & \text{outside }\partial \mathcal{E}(\bm{\mathcal{T}}), \\
        0, & \text{on }\partial \mathcal{E}(\bm{\mathcal{T}}) , \\
        -d(\sigma_{M, xx}, \sigma_{M, yy}), & \text{inside } \partial \mathcal{E}(\bm{\mathcal{T}})
        \end{cases}
    \end{equation}
    where $d(\sigma_{M, xx}, \sigma_{M, yy})$ is the closest point to the boundary \citep{vlassis2021sobolev}.
The Eikonal equation is solved on a regular two-dimensional grid (indexed by $(i,j)$) with a given $\Delta \sigma_{M, xx}$ and $\Delta \sigma_{M, yy}$  using a Fast-Marching solver
\citep{sethian1999fast} which approximates eq. (\ref{eq::Eikonal}) with
\begin{equation}\label{eq::SolEikonal}
    \max \left( \max (D_{ij}^{-x} f, 0)\, , \, \min (D_{ij}^{+x} f, 0)   \right)^{2} + \max \left( \max (D_{ij}^{-y} f, 0)\, , \, \min (D_{ij}^{+y} f, 0)   \right)^{2} = 1
\end{equation}
where the difference operator notation is used, e.g.
\begin{equation}
    \begin{aligned}
        D_{ij}^{-x} f & = (f_{i,j} - f_{i-1,j})/ \Delta \sigma_{M, xx},\\
        D_{ij}^{+x} f & = (f_{i+1,j} - f_{i,j})/ \Delta \sigma_{M, xx}.
    \end{aligned}
\end{equation}
All following results were obtained using the Fast-Marching solver of \cite{furtney2015scikit}.

    Hence, the final regression-based dataset we utilize for training is defined by $\mathcal{D}_{reg} = \left\lbrace \left( ({\sigma}_{M, xx}^{i}, {\sigma}_{M, yy}^{i}, \bm{\mathcal{T}}^{i}),\phi^{i}  \right) \right\rbrace_{i=1}^{N}$ where $\phi^{i}$ represents the solution value of the Eikonal equation at $({\sigma}_{M, xx}^{i}, {\sigma}_{M, yy}^{i})$ for a specific $\bm{\mathcal{T}}^{i}$ value.
    As stated above the solution of the Eikonal equation can be interpreted as a signed distance function which can be used to model the yield function $f$ when only binary values (yield/ not-yield) are available.
    Therefore, utilizing this training dataset allows us to build a predictive tool for the yield function $f$
    \begin{equation}\label{eq::YieldFunPred}
    \begin{aligned}
            \hat{f}   &\equiv  \hat{\phi}({\sigma}_{M, xx}, {\sigma}_{M, yy}, \bm{\mathcal{T}}) \\ 
        \text{with: }&    
            \begin{cases}
   \hat{\phi} \leq 0, & \text{elastic},\\
    \text{else}, & \text{plastic}.
    \end{cases}
    \end{aligned}
    \end{equation}
    
\begin{figure}
\begin{subfigure}[b]{.5\linewidth}
\centering
    \centering
    \includegraphics[scale=0.4]{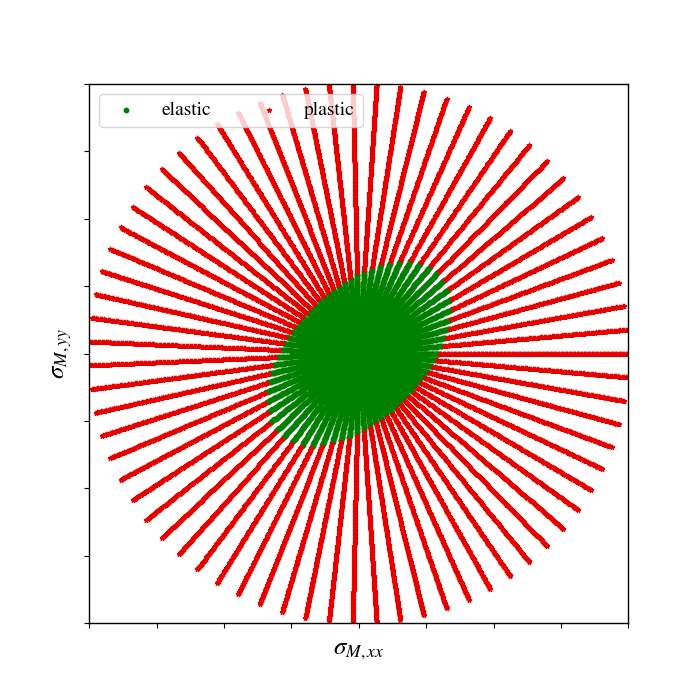}
    \caption{}\label{fig::J2Binary}
\end{subfigure}%
\begin{subfigure}[b]{.5\linewidth}
\centering
    \centering
    \includegraphics[scale=0.4]{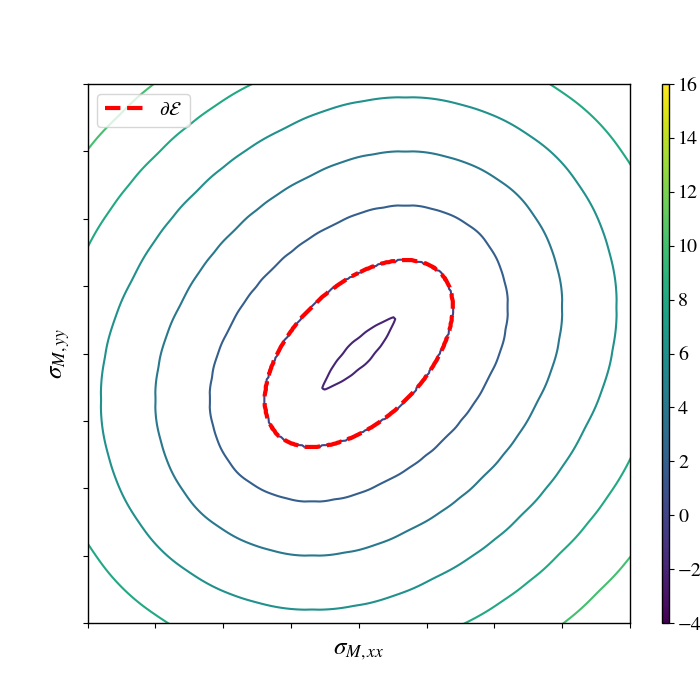}
    \caption{}\label{fig::J2Eikonal}
\end{subfigure}

\begin{subfigure}[b]{.5\linewidth}
\centering
    \centering
    \includegraphics[scale=0.4]{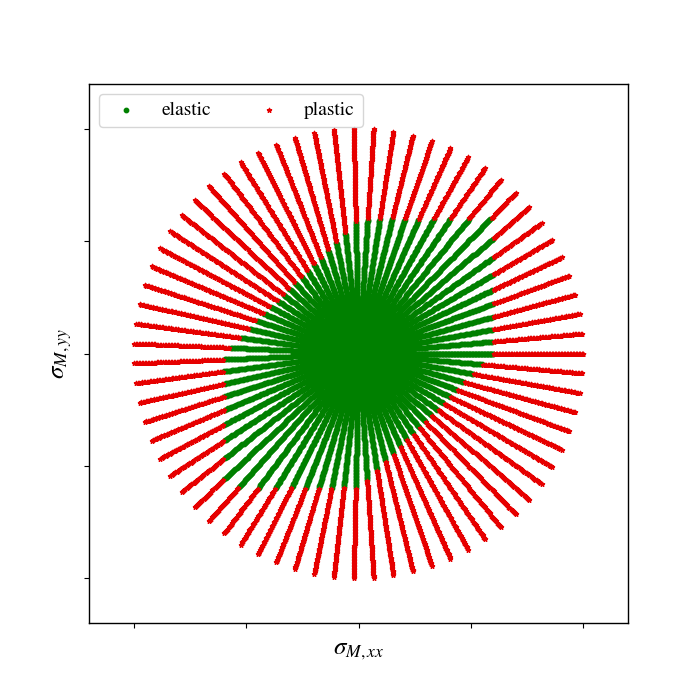}
    \caption{}\label{fig::trescaBinary}
\end{subfigure}%
\begin{subfigure}[b]{.5\linewidth}
\centering
    \centering
    \includegraphics[scale=0.4]{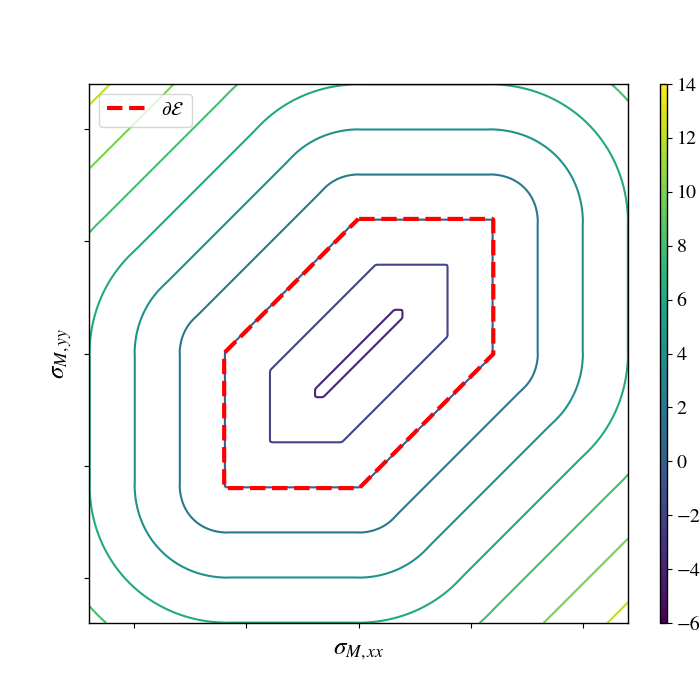}
    \caption{}\label{fig::TrescaEikonal}
\end{subfigure}
    \caption{Initial classification-based data (\protect\subref{fig::J2Binary},\protect\subref{fig::trescaBinary}) and final regression-based data (\protect\subref{fig::J2Eikonal},\protect\subref{fig::TrescaEikonal}) as solution of the Eikonal equation $\abs{\nabla f} = 1$ for the yield surfaces defined by: (\protect\subref{fig::J2Binary},\protect\subref{fig::J2Eikonal}) J2 yield function, (\protect\subref{fig::trescaBinary},\protect\subref{fig::TrescaEikonal}) Tresca yield function. }
    \label{fig:}
\end{figure}

    We test this approach on two commonly used yield functions, first for J2-plasticity 
    \begin{equation}
        f_{J2} = \sqrt{\frac{1}{3} \left( (\sigma_{M, xx} - \sigma_{M, yy})^{2} + \sigma_{M, xx}^{2} + \sigma_{M, yy}^{2}  \right)} - \kappa
    \end{equation}
    and also for the Tresca yield function 
    \begin{equation}
        f_{Tresca} = \frac{1}{2} \max \left( \abs{\sigma_{M, xx} - \sigma_{M, yy}}  \right) - \frac{1}{2} \kappa
    \end{equation}
    in two dimensions where $\kappa$ represents a yield stress value. Consider the binary data shown in Figures \ref{fig::J2Binary} and \ref{fig::trescaBinary} from $72$ loading paths sampled with equidistant angular stepsize.
    The respective solutions of the Eikonal equation are shown in the Figures \ref{fig::J2Eikonal} and \ref{fig::TrescaEikonal}. It can be seen that the presented procedure allows us to quickly reformulate binary macroscopic yield surface data into problems allowing for regression-based machine learning prediction which helps to directly use the trained models in time-integration schemes.

\begin{figure}
\begin{subfigure}[b]{.5\linewidth}
\centering
    \centering
    \includegraphics[scale=0.4]{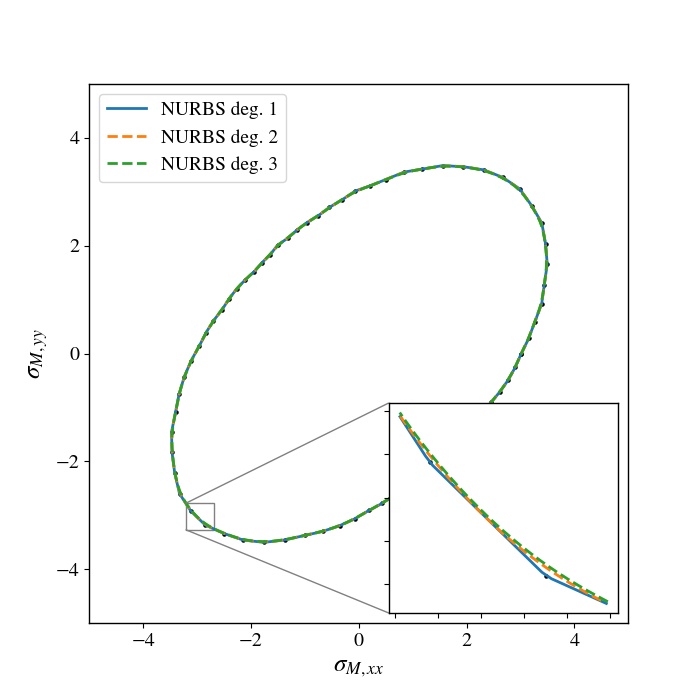}
    \caption{}\label{fig::J2Nurbs}
\end{subfigure}%
\begin{subfigure}[b]{.5\linewidth}
\centering
    \centering
    \includegraphics[scale=0.4]{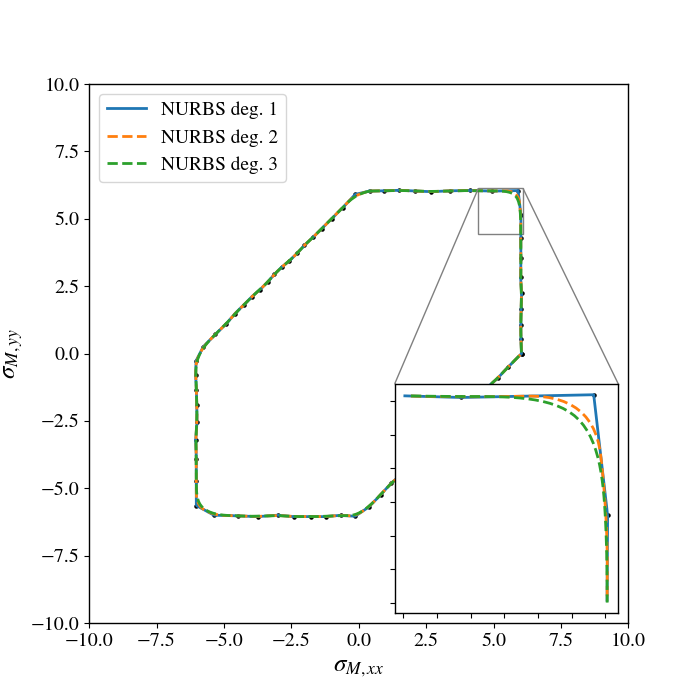}
    \caption{}\label{fig::TrescaNurbs}
\end{subfigure}
    \caption{Exploration of strategies for interpolation of discrete points on yield surface. NURBS curves with degree 1,2,3 for $72$ points in equidistant angular direction for (\protect\subref{fig::J2Nurbs}) J2 yield surface, (\protect\subref{fig::TrescaNurbs}) Tresca yield surface. Black dots represent control points.}
    \label{fig:}
\end{figure}

However, we can incorporate additional constraints about the properties of the yield surfaces into our predictive tool.
In his influential works Drucker \citep{drucker1951more}, \citep{drucker1959definition}
 showed that, based on a stability postulate, the yield surface
must be convex, i.e.
\begin{equation}\label{eq::Convex}
    (\bm{\sigma}_{M}^{\star} - \bm{\sigma}_{M}): \frac{\partial f}{\partial \bm{\sigma}_{M}} \leq 0
\end{equation}
where $\bm{\sigma}^{\star}_{M}$ is an arbitrary macroscopic stress on or inside the yield surface.
\cite{lippmann1970matrixungleichungen} later proved that if a yield function is convex in the three-dimensional space of the principal stresses, then it is also convex in the most general six-dimensional stress space. Hence, a predicted yield function of form of eq. (\ref{eq::YieldFunPred}) which is convex with regards to the arguments $\sigma_{M, xx}$ and $\sigma_{M, yy}$ is also convex in the general stress space.
Hence, we wish to to generate a data-driven yield function that conforms to this property.
This is simplified by the fact that the signed distance function (i.e. the dataset that we are using) of a set is convex when the set is convex \citep{yan2020convexity}.

In the next section we introduce and discuss an approach based on neural networks that allows to train a predictive model whose output is always convex with regards to a subset of its inputs.

    \begin{remark}\label{remark::Nurbs}(Interpolation of discrete points on yield surface)
    Solving eq. (\ref{eq::Eikonal}) using the approach outlined in Section \ref{sec::3} requires an accurate discrete representation of the boundary surface $\partial \mathcal{E}$. Hence, when only a few points of the boundary are known, interpolation methods are needed to fill in the gaps between the control points. However, the choice of interpolation technique has significant influence on the shape of the approximated yield surface. To elaborate on this we evaluate the interpolated yield surface curves using  
    Non-Uniform Rational Basis Splines (NURBS) with basis functions of varying degrees.
    Following \cite{piegl1991nurbs} a NURBS-curve is defined as 
    \begin{equation}
        \bm{C} (u) = \sum_{i=1}^{k}  \frac{ w_{i} N_{i,k}(u)}{\sum_{j=1}^{k}w_{j} N_{j,k}(u)} \bm{P}_{i}
    \end{equation}
    with some weights $w_{i}$, the control points $\bm{P}_{i}$, the curve parametrization $u$ and where $N_{i,k}(u)$ are normalized B-spline basis functions of degree $k$.
    Figures \ref{fig::J2Nurbs} and \ref{fig::TrescaNurbs} show the interpolation of $72$ equidistant points on the yield surface of J2 and Tresca yield function along angular direction using NURBS interpolators with degrees $1,2$ and $3$.  It can be seen that degree $1$ interpolation linearly connects the points along the curve, whereas degrees of higher order smooth the yield surfaces out. For this reason NURBS with degree $1$ will in the following be used for interpolating the yield surfaces as a preparation for solving the discrete form of the Eikonal equation in eq. (\ref{eq::SolEikonal}).
    \end{remark}
    
    \begin{remark}(Normalization of dataset)
    % https://www.tandfonline.com/doi/pdf/10.3846/1611-1699.2008.9.79-86
    In order to speed up the training process and to make training more reliable \citep{goodfellow2016deep}, we employ Min-max feature scaling to bring all values of each feature into the range $[0,1]$. For the measure $\bm{X}$, the scaling process reads
    \begin{equation}
        \overline{X}_{ij} = \frac{X_{ij} - \min_{j} X_{ij}}{\max_{j} X_{ij} - \min_{j} X_{ij}} 
    \end{equation}
    where $\overline{X}_{ij}$ is the scaled output.
    \end{remark}
    
\section{Machine learning formulation for smooth and convex yield functions}\label{sec::4}
In this section we give a brief overview of the general neural network formulation and then specify the particular architectures for neural networks that achieve convexity with regards to the input.
For this consider the data set 
$\mathcal{D}_{reg} = \left\lbrace \left( \bm{y}^{i} , \phi^{i}  \right) \right\rbrace_{i=1}^{N}$
where we used the simplification $\bm{y}^{i} = ({\sigma}^{i}_{M, xx}, {\sigma}^{i}_{M, yy}, \bm{\mathcal{T}}^{i}) $ consisting of $N$ samples. We furthermore define $\bm{y} = [\bm{y}^{c}, \bm{y}^{nc}]$ where $\bm{y}^{c, i}= [{\sigma}_{M, xx}^{i}, {\sigma}^{i}_{M, yy}]$ and $\bm{y}^{nc,i }= \bm{\mathcal{T}}^{i}$. Here $\bm{y}^{c}$ and $\bm{y}^{nc}$ represent the subsets of $\bm{y}$ that the output is required to be convex to and not necessarily required to be convex to, respectively.

\subsection{Neural networks}
\begin{figure}
\begin{subfigure}[b]{.25\linewidth}
\centering
    \centering
    \includegraphics[scale=0.25]{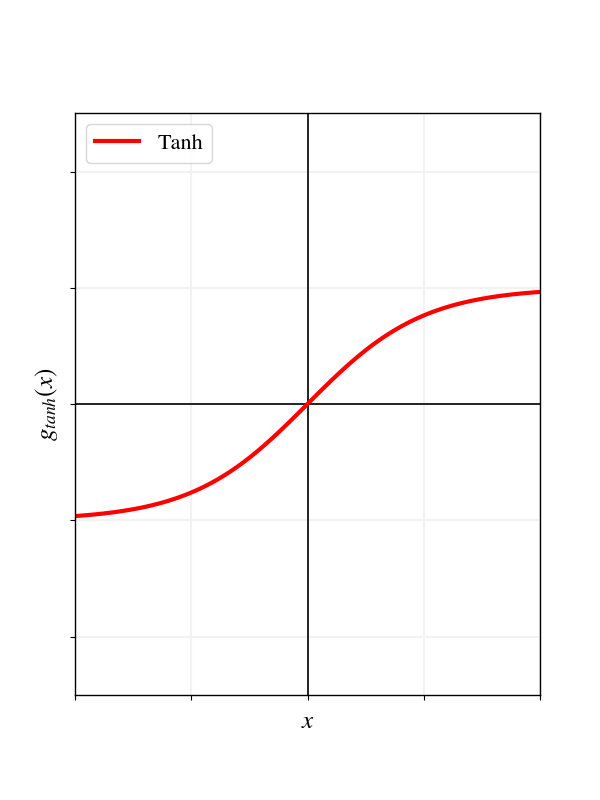}
    \caption{}\label{fig::actTanh}
\end{subfigure}%
\begin{subfigure}[b]{.25\linewidth}
\centering
    \centering
    \includegraphics[scale=0.25]{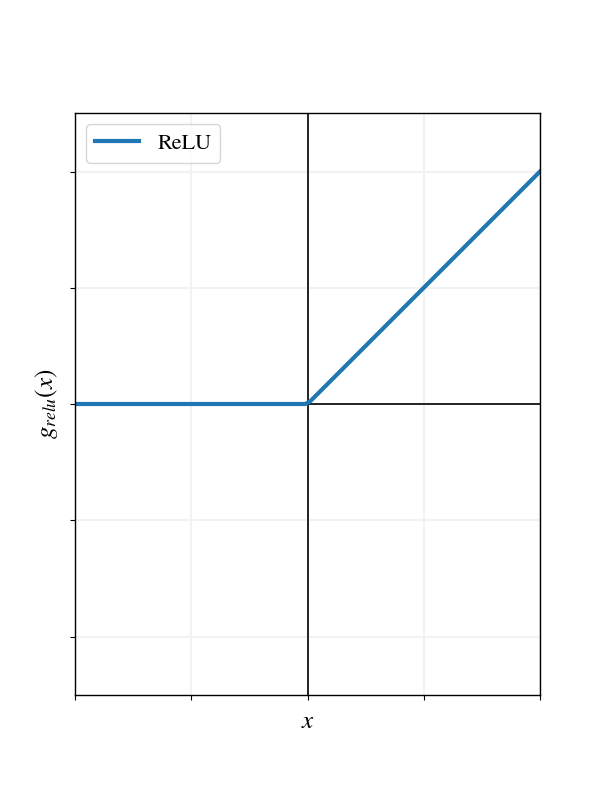}
    \caption{}\label{fig::actRelu}
\end{subfigure}%
\begin{subfigure}[b]{.25\linewidth}
\centering
    \centering
    \includegraphics[scale=0.25]{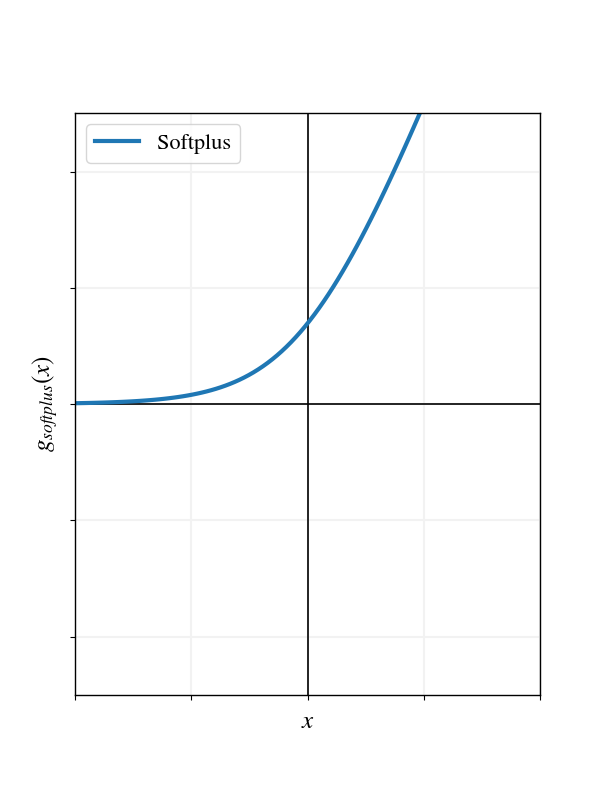}
    \caption{}\label{fig::actsoft}
\end{subfigure}%
\begin{subfigure}[b]{.25\linewidth}
\centering
    \centering
    \includegraphics[scale=0.25]{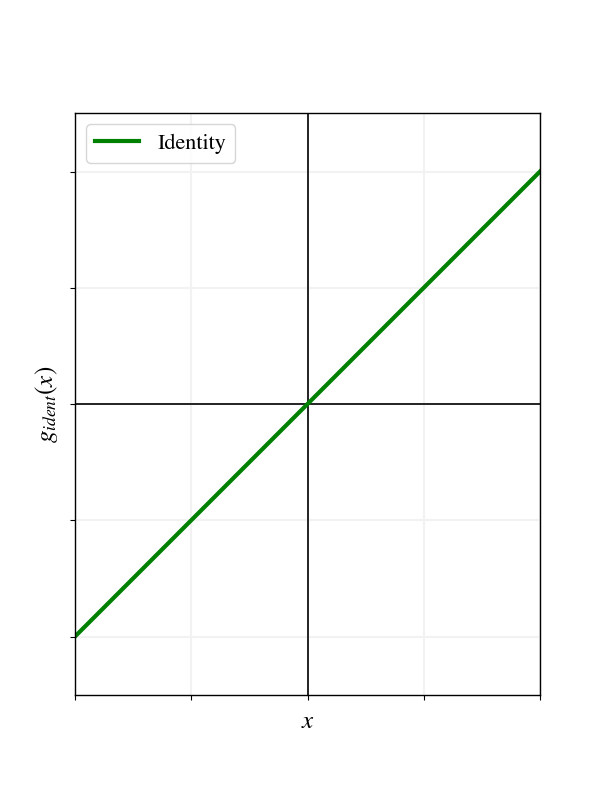}
    \caption{}\label{fig::actIdent}
\end{subfigure}%

    \caption{Three activation function commonly found in neural networks: (\protect\subref{fig::actTanh}) Tanh activation function, (\protect\subref{fig::actRelu}) Rectified Linear Unit (ReLU), 
    (\protect\subref{fig::actsoft}) Softplus activation function
    and  (\protect\subref{fig::actIdent}) Identity function.}
    \label{fig:}
\end{figure}

A typical feedforward neural network comprises of one input layer, $n_{D}-1$ hidden layers and one output layer. 
Let each of the hidden layers consist of $n_{k}$ neurons and have the output $\bm{z}_{k} \in \mathbb{R}^{n_{k}}$ where  $k=1, \ldots, n_{D}$. Generally, the output of the $k^{\text{th}}$ layer is obtained by the transformation
\begin{equation}\label{eq::NNO}
   \bm{z}_{k} = \mathcal{L} (\bm{z}_{k-1})  = g_{k}\left(\bm{W}_{k} \bm{z}_{k-1} + \bm{b}_{k} \right)
\end{equation}
with $\bm{z}_{0}=\bm{y}$, $\bm{z}_{n_{D}} = \phi$ and
where $\bm{W}_{k} $, $\bm{b}_{k}$ and $g_{k}$ are the weights, biases and activation function of the $k^{\text{th}}$ layer. Common choices of the activation functions in the hidden layers include Tanh, and Rectified Linear Unit (ReLU) (see Figures \ref{fig::actTanh} and \ref{fig::actRelu}) which are given by
\begin{equation}
    \begin{aligned}
        g_{tanh}(x) &= \frac{2}{1+e^{-2x}} - 1\\
        g_{relu}(x) &= \max (0,x).
    \end{aligned}
\end{equation}
A smoothed version of ReLU (i.e. continuously differentiable) is Softplus (see Figure \ref{fig::actsoft})
\begin{equation}
    g_{softplus}(x) = \log (1 + e^{x}).
\end{equation}
For regression problems, the activation function of the output layer is chosen as the identity function $g_{ident}(x) = x $, see Figure \ref{fig::actIdent}.
Hence, an input value $\bm{y}$ into the network yields the output $\hat{\phi}$ with
\begin{equation}
    \hat{\phi}(\bm{y}) = (\mathcal{L}_{k} \circ \mathcal{L}_{k-1}\circ \cdots \circ \mathcal{L}_{1})(\bm{y})
\end{equation}
where $\circ$ is a composition operator. The unknown and trainable parameters are given by the set $\bm{\Theta} = \lbrace \bm{W}^{k}, \bm{b}^{k}  \rbrace_{k=1}^{n_{D}}$.
The optimal values $\bm{\Theta}^{\star}$ need to be obtained by defining an optimization problem over a loss function $L(\bm{\Theta} )$
\begin{equation}\label{eq::NNOpti}
    \bm{\Theta}^{\star} = \argmin_{\bm{\Theta}} L(\bm{\Theta} ).
\end{equation}
In the context of this work we use the mean-squared error as the loss function
\begin{equation}
    L(\bm{\Theta} ) = \text{ } \frac{1}{N} \sum\limits_{i=1}\limits^{N} \norm{ \phi^{i}-\hat{\phi}^{i}}_{2}^{2} .
    \end{equation}
Due to the complexity of the optimization problem eq. (\ref{eq::NNOpti}) the optimal parameter set is commonly approximated in an iterative manner by a stochastic gradient descent (SGD) algorithm such as the SGD with momentum \citep{sutskever2013importance} or variants such as ADAM \citep{kingma2014adam}.
More general information on neural networks can be found in relevant textbooks, e.g. \cite{goodfellow2016deep}. 

If one was to use this (most commonly employed) neural network architecture, the convexity of the data-driven yield function can not be guaranteed since the presented neural network output is not necessarily convex with regards to the inputs. 
To overcome issues like this, \cite{amos2017input} introduced an approach termed input convex neural networks that offer a ML-based predictive tool that is convex by design. This is not merely a penalization of convexity by adding one more term in the loss function, but a strict enforcement of convexity through a specialized architecture.

\subsection{(Partially) input convex neural networks}
\begin{figure}[h!]
    \centering
    \includegraphics[scale=0.5]{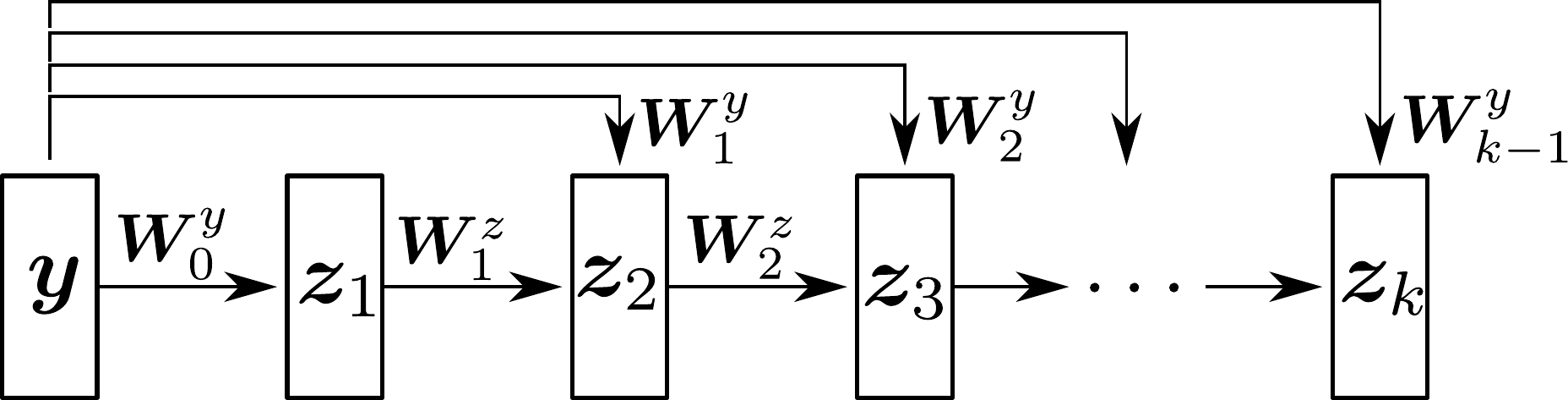}
    \caption{Visual representation of the network architecture of input convex neural networks.}
    \label{fig:icnn}
\end{figure}

In order to ensure that the output of the neural network is convex with regards to all input dimensions of $\bm{y}$ while still offering an expressive neural network, \cite{amos2017input} rewrote the standard update formula of eq. (\ref{eq::NNO}) to 
\begin{equation}
    \bm{z}_{i+1} = g_{i} (\bm{W}_{i}^{z} \bm{z}_{i} + \bm{W}_{i}^{y} \bm{y} + \bm{b}_{i}),
\end{equation}
where $\bm{W}_{0}^{z} = \bm{0}$ and $\bm{z}_{0} = \bm{0}$. The set of trainable parameters is given by $\lbrace \bm{W}_{1:k-1}^{z}, \bm{W}_{0:k-1}^{y}, \bm{b}_{0:k-1} \rbrace$.
The resulting network architecture is schematized in Figure \ref{fig:icnn}.
It can be seen that this formulation includes "passthrough" layers, i.e. the input $\bm{y}$ is directly connected to the hidden and output layers. The output of this network is convex with regards to the inputs if all weights $\lbrace \bm{W}_{i}^{z} \rbrace_{i=1}^{k-1}$ are non-negative and the activation functions $g_{i}$ are non-decreasing and convex. For a proof refer to \cite{boyd2004convex} (3.2).
An example of a non-deceasing and convex activation function is ReLU as visualized in Figure \ref{fig::actRelu} which is also the function of choice of \cite{amos2017input}. 
Later, \cite{chen2020input} replaced ReLU by Softplus activation functions (Figure \ref{fig::actsoft}).

\begin{figure}[h!]
    \centering
    \includegraphics[scale=0.5]{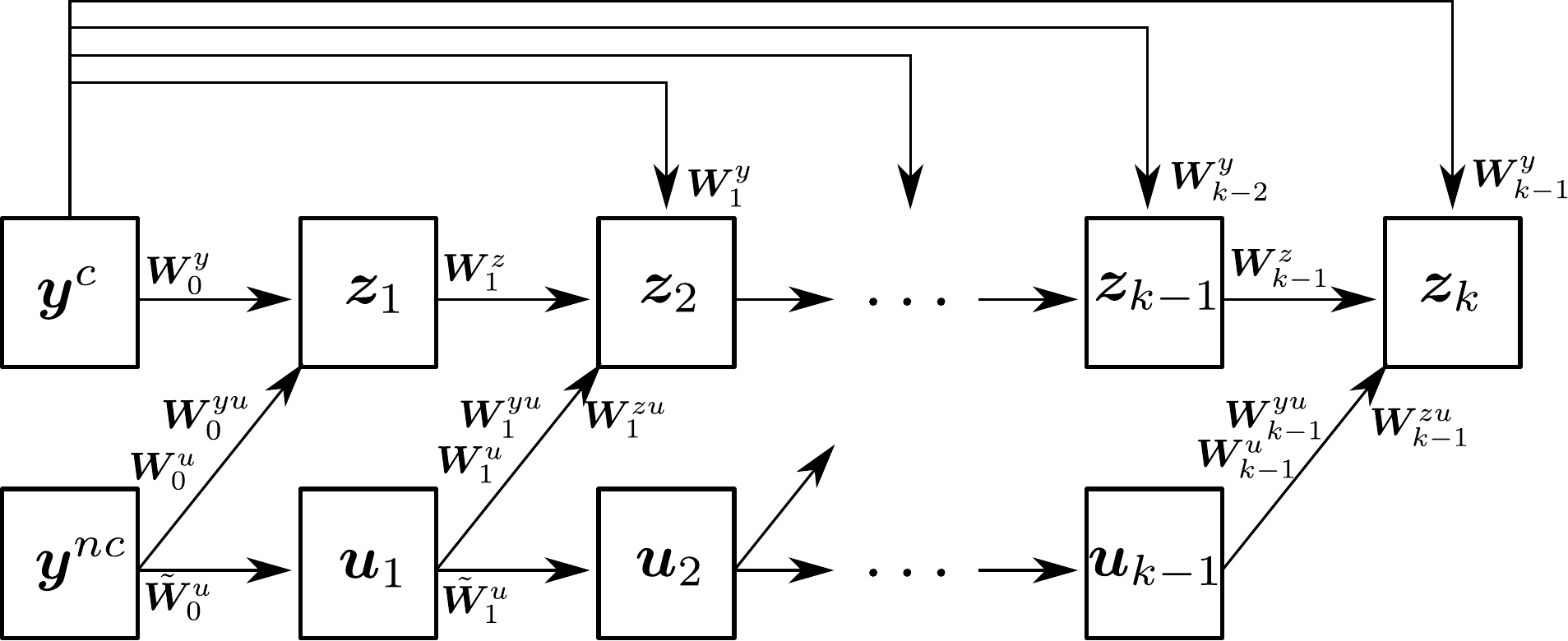}
    \caption{Visual representation of the network architecture of partially input convex neural networks}
    \label{fig:picnn}
\end{figure}

Conveniently, in this work we do not require the neural network output to be convex to all input dimensions $\bm{y}$ but only to the stress subset $\bm{y}^{c}$. In this case the update formula reads
\begin{equation}
\begin{aligned}
        \bm{u}_{i+1} &= \tilde{g}_{i} (\tilde{\bm{W}}_{i}^{u} \bm{u}_{i}  + \tilde{\bm{b}}_{i}) \\
        \bm{z}_{i+1} &= g_{i} \left(\bm{W}_{i}^{z} (\bm{z}_{i} \circ [\bm{W}_{i}^{zu} \bm{u}_{i} + \bm{b}_{i}^{z}]) + \bm{W}_{i}^{y} (\bm{y}^{c} \circ [\bm{W}_{i}^{yu} \bm{u}_{i} + \bm{b}_{i}^{y}]) + \bm{W}_{i}^{u} \bm{u}_{i} + \bm{b}_{i}  \right)
\end{aligned}
\end{equation}
where $\bm{W}_{0}^{z} = \bm{0}$, $\bm{z}_{0} = \bm{0}$ and $\bm{u}_{0} = \bm{y}^{nc}$. The set of trainable parameters is given by $\lbrace \bm{W}_{1:k-1}^{z}, \bm{W}_{0:k-1}^{y}, \tilde{\bm{W}}_{0:k-1}^{u}, \bm{W}_{0:k-1}^{zu},  \bm{W}_{0:k-1}^{yu}, \tilde{\bm{b}}_{0:k-1}, \bm{b}_{0:k-1} \rbrace$. An overview of the network architecture is given in Figure \ref{fig:picnn}. 
Similarly to the previous case, the output is convex with respect to the inputs $\bm{y}^{c}$ if all weights $\lbrace \bm{W}_{i}^{z} \rbrace_{i=1}^{k-1}$ are non-negative and the activation functions $g_{i}$ are non-decreasing and convex. 
Using this approach we are able to train a predictive model for the yield function that ensures its convexity with regards to the stresses but does not enforce convexity with regards to the texture parameters. 

        \begin{remark}(Representation of vertices and derivatives at non-differentiable points)
        With traditional phenomenological models for yield functions vertices are either deliberately smoothed out to allow for their utilization in time-integration frameworks or the iteration loops need to be tediously adjusted for the edge cases which is not trivial, see for example \cite{peric1999new}. Using ICNNs however we potentially are able to accurately reproduce the vertex as well as obtain derivative values at these points of interest, see Appendix \ref{sec::AppendixVertices}. This allows us to directly incorporate any trained model (with or without vertices present) in a standard time-integration loop.
            \end{remark}

    \begin{remark}(Initialization of weights)
    In order to allow for activation variances and back-propagated gradients variance when propagating inputs forward and outputs backward through the network the weights of the network are initialized using Glorot's uniform distribution \citep{glorot2010understanding} which reads
    \begin{equation}
        W \sim \mathcal{U} \left[ -\sqrt{\frac{6}{n_{k} + n_{k+1}}}, \sqrt{\frac{6}{n_{k} + n_{k+1}}} \right]
    \end{equation}
    where $n_{k}$ and $n_{k+1}$ are the number of input and output units in the weight tensor respectively. 
    Following \cite{glorot2010understanding} we furthermore initially set the bias vectors to be zero.
    \end{remark}

\section{Numerical results}\label{sec::5}
In the following the proposed approach is tested.
The pICNN formulation was implemented in
Pytorch \citep{NEURIPS2019_9015}\footnote{The Python code for pICNN developed for this work can be made available under reasonable request.} and the network parameters were optimized using the Adam optimizer \citep{kingma2014adam} with a constant learning rate of $1e-4$. All of the considered pICNNs use ReLU activation functions and consist of 4 hidden layers which results in $\approx 7,200$ trainable parameters. The choices of number of hidden layers as well as the learning rate are not the results of any hyperparameter study. They do not represent any special network setup to the best
of the authors knowledge, they were simply the first hyperparameters that were tested. We highlight this point, as this paper attempts a proof of concept where the goal is not to train the best possible fit, but obtain a reasonable predictive tool with capabilities beyond those of any existing model. Hence, as long as the learning rate is sufficiently low and the network is sufficiently expressive we expect similar results to the ones presented in the following. 

For the crystallographic texture, we consider nine different spread values in the interval $\theta_{m}=5^{\circ}$ to $\theta_{m}=25^{\circ}$ with a step size of $2.5^{\circ}$. For each of the resulting mean values from the corresponding CP simulations and for each value of spread, %({\color{red}{JF: where to put this?}}) 
yield surfaces which are represented by $72$ evenly spaced values in angular direction were obtained and linearly interpolated in between. The Eikonal equation is solved on a uniform grid of $[-500,500]\times [-500,500]$ nodes with $301$ evenly spaced inputs in each direction for each $\theta_{m}$. Hence the full dataset $\mathcal{D}_{reg} = \left\lbrace \left( ({\sigma}_{M, xx}^{i}, {\sigma}_{M, yy}^{i}, \theta_{m){j}}^{i}),\phi^{i}  \right) \right\rbrace_{i=1}^{N}$
consists of $N \approx 815,000$ samples with an input dimension of $3$ and a one-dimensional output.  In order to reduce memory usage we employ a batch-size of $N/250$. We furthermore denote the stress-yield dataset for a specific $\theta^{j}_{m}$ with $\mathcal{D}_{reg, \theta^{j}_{m}} = \left\lbrace \left( ({\sigma}_{M, xx}^{i}, {\sigma}_{M, yy}^{i}),\phi^{i}  \right) \right\rbrace_{i=1}^{N_{\theta^{j}_{m}}}$ with $j=1,\ldots, 9$ where $N_{\theta^{j}_{m}}$ is the size of this subset.

In the following we study how the presented approach performs for In-Sample and Out-Of-Sample predictions. In-Sample predictions means that we test the performance of the predictive tool on samples that were inside the training set whereas Out-Of-Sample prediction tests on unseen input data. 

\subsection{In-Sample prediction}
\begin{figure}
\begin{subfigure}[b]{.5\linewidth}
\centering
    \centering
    \includegraphics[scale=0.4]{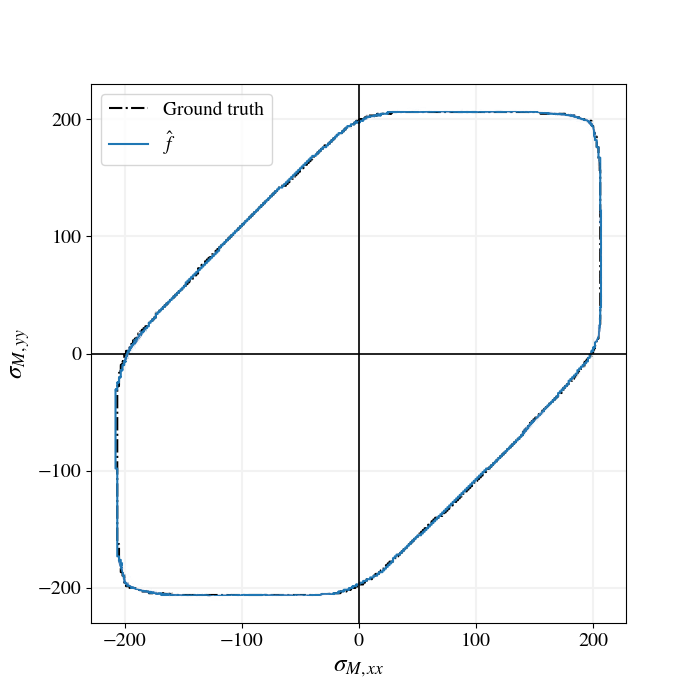}
    \caption{$\theta_{m}=5^{\circ}$}\label{fig::}
\end{subfigure}%
\begin{subfigure}[b]{.5\linewidth}
\centering
    \centering
    \includegraphics[scale=0.4]{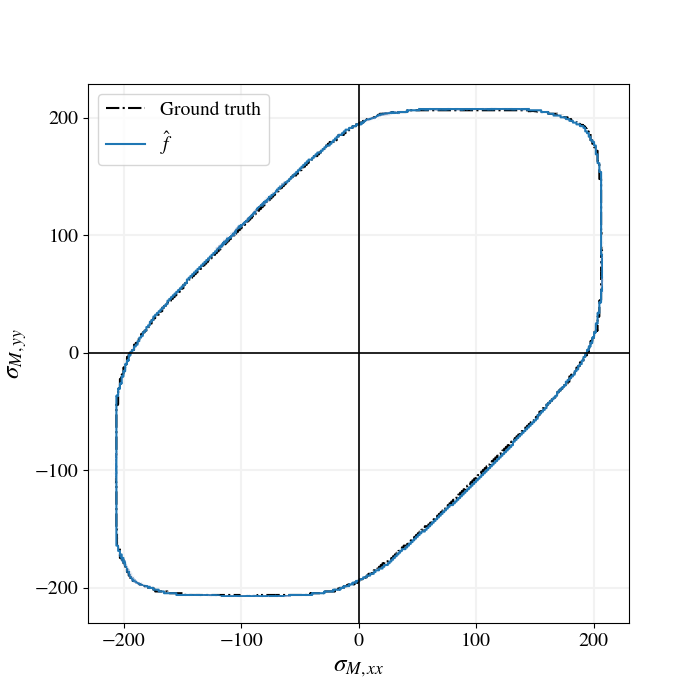}
    \caption{$\theta_{m}=7.5^{\circ}$}\label{fig::}
\end{subfigure}

\begin{subfigure}[b]{.5\linewidth}
\centering
    \centering
    \includegraphics[scale=0.4]{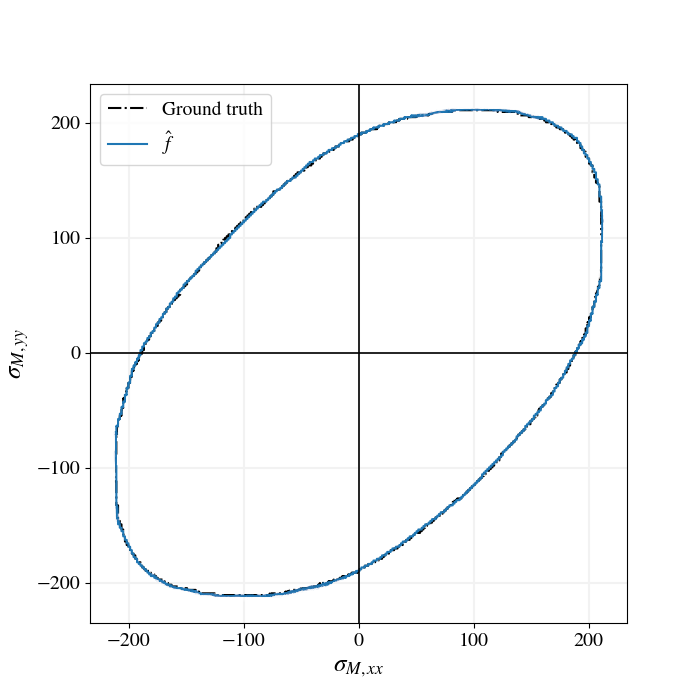}
    \caption{$\theta_{m}=20^{\circ}$}\label{fig::}
\end{subfigure}%
\begin{subfigure}[b]{.5\linewidth}
\centering
    \centering
    \includegraphics[scale=0.4]{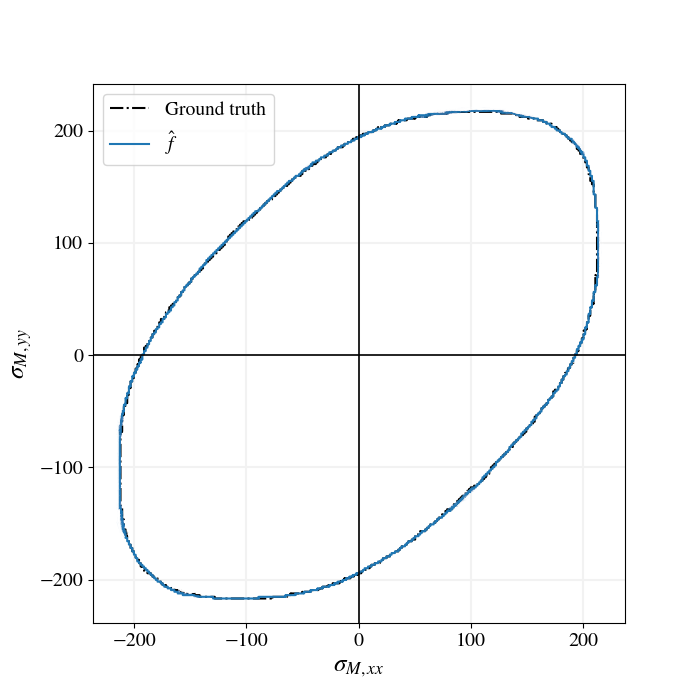}
    \caption{$\theta_{m}=25^{\circ}$}\label{fig::}
\end{subfigure}%
    \caption{In-sample predictions of the mean yield compared to the ground truth means.}
    \label{fig:InSamplePred}
\end{figure}

\begin{figure}[b!]
    \centering
    \includegraphics[scale=0.4]{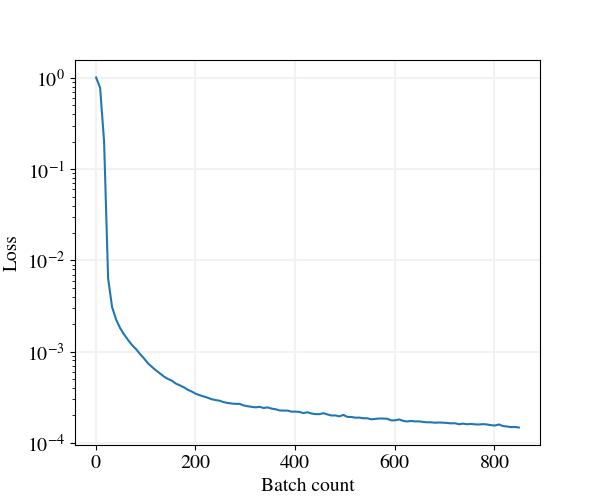}
    \caption{Training with pICNN over whole dataset. Training loss over batch count.}
    \label{fig::InSampleLoss}
\end{figure}

First we are interested in how well the pICNN prediction works for In-Sample predictions. For this we compare the interpolated yield surface output of the neural network with the ground truth (
Figure \ref{fig:InSamplePred}). It can be seen that the predicted yield surfaces for the two extreme cases of the spread-values $\theta_{m} = \lbrace 5^{\circ}, 25^{\circ} \rbrace $ as well as for two non-extreme input values $\theta_{m} = \lbrace 7.5^{\circ}, 20^{\circ} \rbrace$ very accurately coincide with the ground truth. Furthermore due to the nature of the ICNN the predicted output surfaces are necessarily convex.  
The training loss over 850 batches is shown in
Figure \ref{fig::InSampleLoss}. We can see that the final loss value is sufficiently small, around $10,000$-times smaller than the initial loss value. On a conventional laptop (Nvidia Quadro P520, 16GB) employing CUDA \citep{cuda} the training took $\approx 60$ minutes. \\
Furthermore we can notice that
we do not seem to overfit which is possibly due to the amount of data we are able to synthetically create by solving the Eikonal equation.

\subsection{Out-Of-Sample prediction through Leave-One-Out cross-validation}
Next, we study how well the ICNN generalizes on unseen data. For this we employ
Leave-One-Out Cross-Validation (LOOCV) as a special case of the general k-fold cross-validation \cite{fuhg2020state}. We base our cross-validation on leaving out all data represented by a specific spread value, i.e.
\begin{equation}
    \mathcal{D}_{-j} = \mathcal{D}_{reg} \setminus \mathcal{D}_{reg, \theta^{j}_{m}} \,\,\,, \qquad j=1,\ldots,9.
\end{equation}
This allows us to use the dataset $\mathcal{D}_{-j}$ for training while testing the performances of the trained model using $\mathcal{D}_{reg, \theta^{j}_{m}} $. Here, we can distinguish between testing data that only requires interpolation, i.e. leaving out data obtained from $\theta_{m} = \lbrace 7.5^{\circ}, \ldots, 22.5^{\circ} \rbrace$ and cases where extrapolation beyond the training data is needed, i.e. leaving out  $\theta_{m} = \lbrace 5^{\circ}, 25^{\circ} \rbrace$.
\subsubsection{Interpolation}
\begin{figure}
\begin{subfigure}[b]{.5\linewidth}
\centering
    \centering
    \includegraphics[scale=0.4]{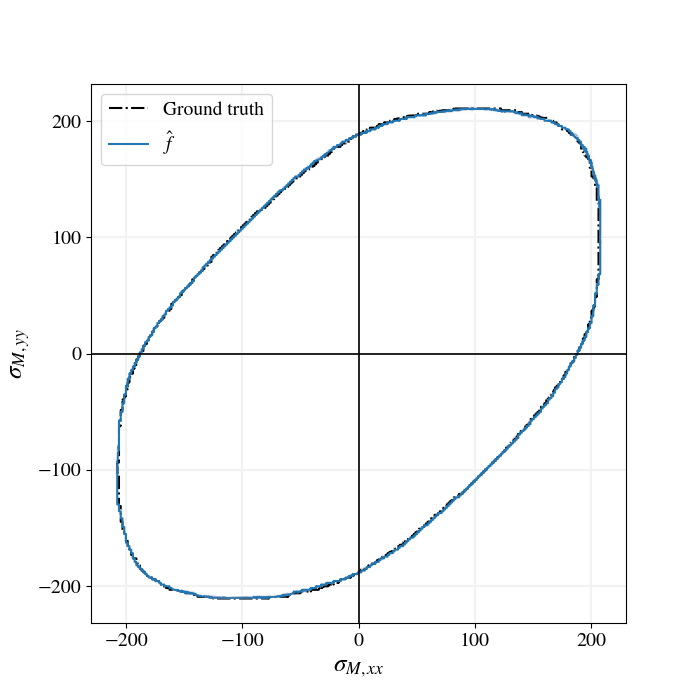}
    \caption{$\theta_{m}=12.5^{\circ}$}\label{fig::}
\end{subfigure}%
\begin{subfigure}[b]{.5\linewidth}
\centering
    \centering
    \includegraphics[scale=0.4]{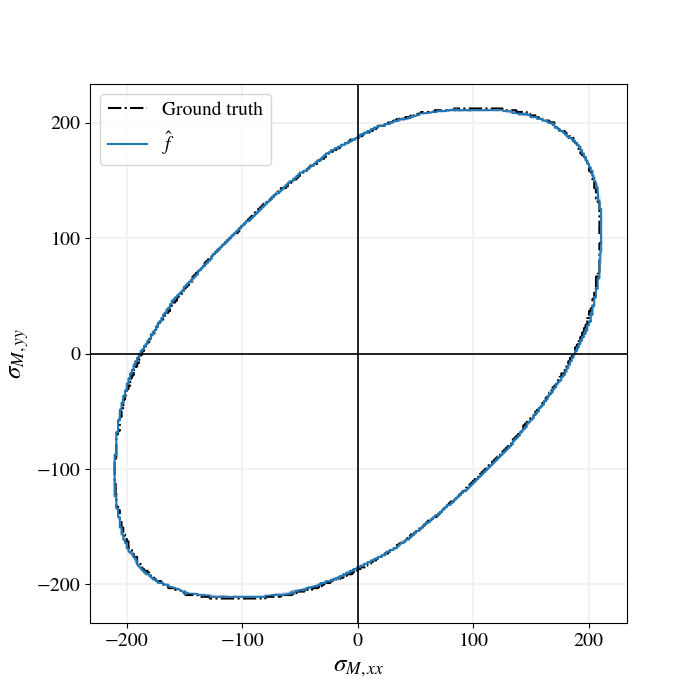}
    \caption{$\theta_{m}=17.5^{\circ}$}\label{fig::}
\end{subfigure}%
    \caption{Out-Of-Sample prediction: Ground truth and predicted mean for LOO-samples requiring interpolation.}
    \label{fig:LOOInterpolation}
\end{figure}

\begin{figure}
\begin{subfigure}[b]{.5\linewidth}
\centering
    \centering
    \includegraphics[scale=0.4]{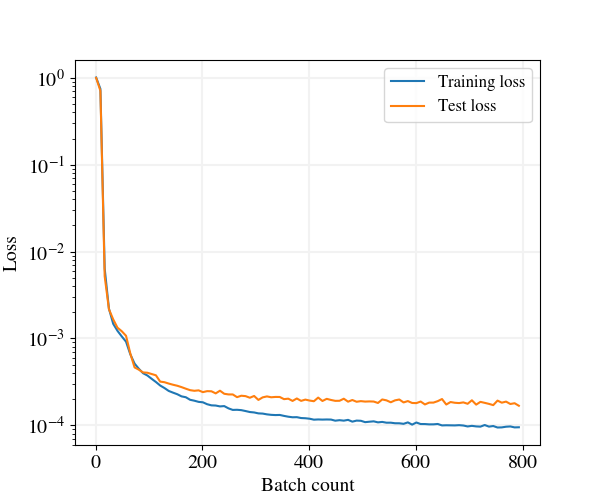}
    \caption{$\theta_{m}=12.5^{\circ}$}\label{fig::}
\end{subfigure}%
\begin{subfigure}[b]{.5\linewidth}
\centering
    \centering
    \includegraphics[scale=0.4]{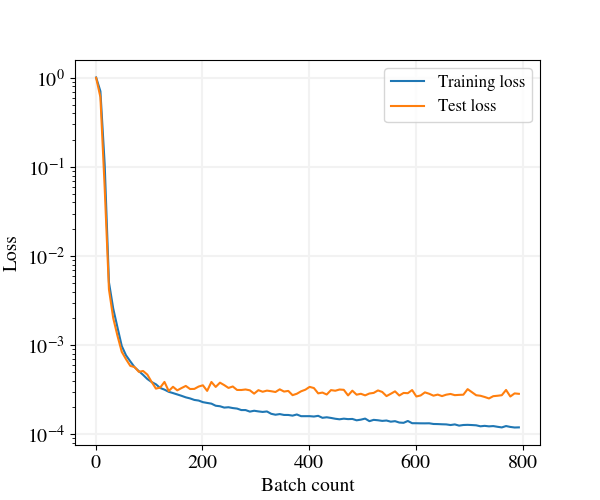}
    \caption{$\theta_{m}=17.5^{\circ}$}\label{fig::}
\end{subfigure}%
    \caption{Training and testing loss for LOO-samples requiring interpolation.}
    \label{fig:LOOInterpolationLoss}
\end{figure}

To investigate the generalization capability of the trained networks for values requiring interpolation we compare the ground truth of $12.5^{\circ}$ and $17.5^{\circ}$ to their predicted counterparts (which are not part of the training set) in Figure \ref{fig:LOOInterpolation}. From this visual comparison we can see that the generalization for interpolated values is surprisingly accurate, i.e. the shapes are reproduced proficiently well.
The training and testing loss over $800$ batches for these two cases are displayed in
Figure \ref{fig:LOOInterpolationLoss}. We can see that even though the ICNN has never seen the testing data, over the course of the training it is able to reduce the error by a factor of around $4,000$ compared to the initial loss. This proves the correlation in the data and that the neural network generalizes very proficiently.

\subsubsection{Extrapolation}
\begin{figure}
\begin{subfigure}[b]{.5\linewidth}
\centering
    \centering
    \includegraphics[scale=0.4]{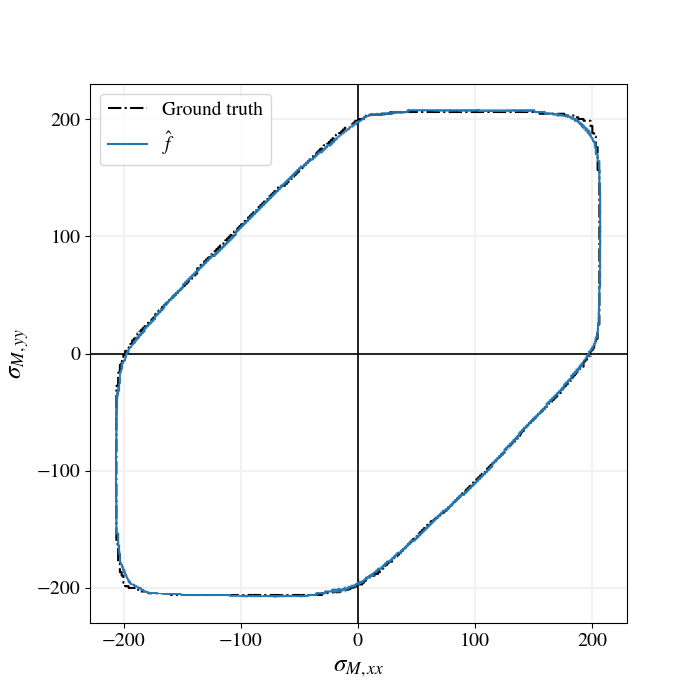}
    \caption{$\theta_{m}=5.0^{\circ}$}\label{fig::}
\end{subfigure}%
\begin{subfigure}[b]{.5\linewidth}
\centering
    \centering
    \includegraphics[scale=0.4]{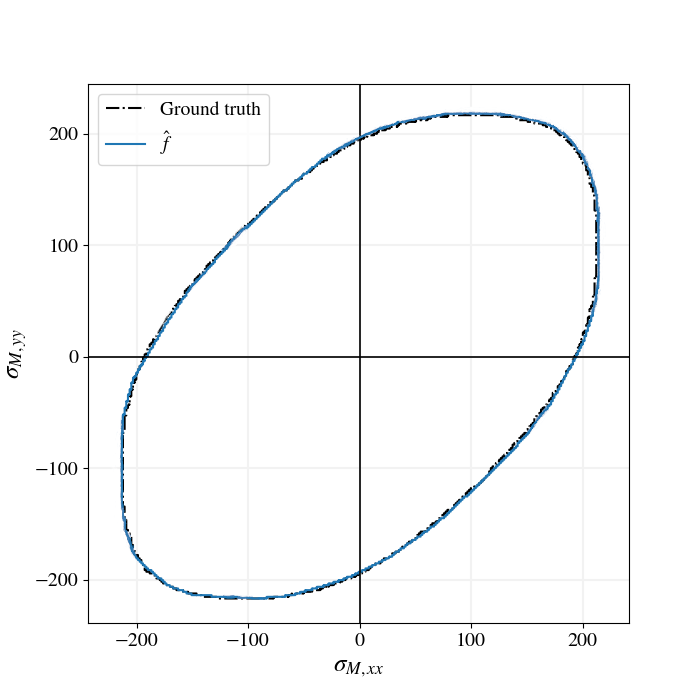}
    \caption{$\theta_{m}=25.0^{\circ}$}\label{fig::}
\end{subfigure}%
    \caption{Out-Of-Sample prediction: Ground truth and predicted mean for LOO-samples requiring extrapolation.}
    \label{fig:LOOExtrapolation}
\end{figure}

Going one step further we use LOOCV to study the performances on the trained network for the two edge cases which require extrapolation when left out of the training dataset, i.e. $\theta_{m}=5^{\circ}$ and $\theta_{m}=25^{\circ}$ respectively. The ground truth yield surfaces and the predictions are visually compared in  
Figure \ref{fig:LOOExtrapolation}. It can be seen that we are able to accurately capture the shapes of the yield surfaces surprisingly well even though the vertices for $\theta_{m}=5^{\circ}$ could not be reproduced exactly. More promisingly, from this visual comparison we can see that the absolute errors are still $<10$ MPa.   

\begin{figure}
\begin{subfigure}[b]{.5\linewidth}
\centering
    \centering
    \includegraphics[scale=0.4]{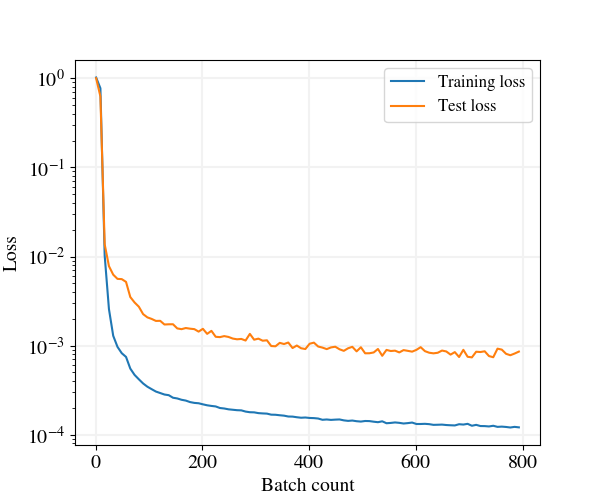}
    \caption{$\theta_{m}=5.0^{\circ}$}\label{fig::}
\end{subfigure}%
\begin{subfigure}[b]{.5\linewidth}
\centering
    \centering
    \includegraphics[scale=0.4]{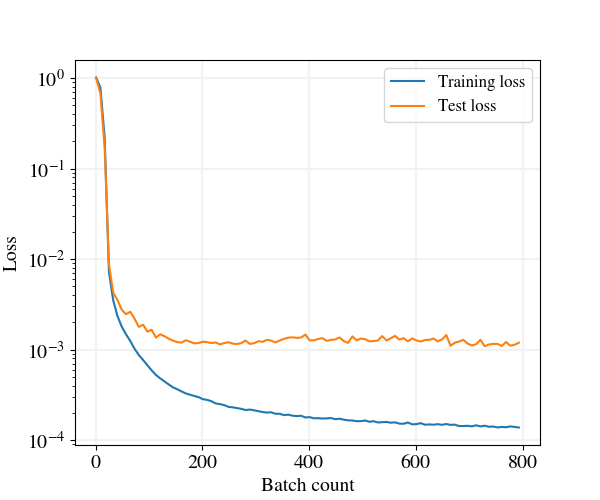}
    \caption{$\theta_{m}=25.0^{\circ}$}\label{fig::}
\end{subfigure}%
    \caption{Training and testing loss for LOO-samples requiring extrapolation.}
    \label{fig:LOOExtrapolationLoss}
\end{figure}

In Figure \ref{fig:LOOExtrapolationLoss} the  training and testing loss over $800$ batches for these two edge cases is plotted. It is noticeable that the testing error reduces significantly even though the neural network has not been trained with the data and the data is outside of the parameter domain.
For both cases we are able to reduce the error by a factor of around $1,000$ compared to the initial loss.
This allows us to reach three conclusions, (1) the proposed neural network generalizes well, (2) the spread is significantly correlated with the change of the yield surfaces, (3) the yield surfaces do not appear to change abruptly outside of the training domain allowing for reliable extrapolation.
However, it should be noted that in comparison to the interpolated cases (Figure \ref{fig:LOOInterpolationLoss}) the network has a higher loss when extrapolation is necessary.

\section{Discussion and Outlook}\label{sec::6}
In this work we present a framework for predicting texture-dependent smooth and convex macroscopic yield functions from data of crystal-plasticity simulations of polycrystals. 
Instead of the traditional phenomenological approach we utilize machine learning to create a predictive tool for the yield function.
In order to ensure the convexity of the trained surface we rely on partially input convex neural networks that ensure that the output of the neural network is convex with regards to the stress inputs.

We test the framework on a dataset using the spread of orientation distribution as an additional input to the network. The network is able to accurately reproduce In-Sample inputs and also shows surprising accuracy for generalization which was highlighted using Out-Of-Sample inter- and extrapolation examples.
Overall, as a proof-of-concept, we show that the presented framework is able to directly incorporate texture-parameters into the fitting and prediction of macroscopic yield surfaces. Additionally, for the first time we showcase that ML tools can enable data-driven yield functions that can respect the Drucker convexity postulate, and can be potentially implemented in a straightforward fashion in standard time-integration schemes. 

Future work aims to extend and optimize this framework. The primary limitation of the current iteration lies in the computation time of simulations. This is influenced primarily by the mesh size (i.e., number of elements) and the number of load vectors considered. The degree to which either of these factors influence the shape of the yield surface and the error in prediction deserve a full-fledged study. This will shed light on the possibility to employ adaptive sampling techniques \citep{fuhg2020state}, or multi-fidelity adaptive sampling \citep{fuhg2019adaptive} (when yield functions of different fidelity can be obtained) to optimize computation time to provide predictions within a known, acceptable error limit. The framework will likewise be employed to consider more complex textures and material descriptions, and will be embedded into a structural finite-element code to help bridge microstructure-component length scales.

%In future works we aim to extend this framework to consider additional texture parameters and embed it into a structural Finite-Element code. Furthermore, in order to possibly reduce the number of simulations needed to generate the database, adaptive sampling techniques \citep{fuhg2020state} might be explored. 

\section*{Acknowledgements}

LVW and MPK were funded or partially funded by Air Force Research Lab grant FA8650-20-1-5203. NB acknowledges startup support from Cornell University. We would like to thank Dr. Romain Quey of {\'E}cole des Mines de Saint-{\'E}tienne for development of Neper capabilities necessary for this work.

\appendix
\section*{Appendices}
\addcontentsline{toc}{section}{Appendices}
\renewcommand{\thesubsection}{\Alph{subsection}}
\subsection{Fitting and smoothing of vertices with input convex neural networks}\label{sec::AppendixVertices}
\begin{figure}[ht!]
\begin{subfigure}[b]{.5\linewidth}
\centering
    \centering
    \includegraphics[scale=0.35]{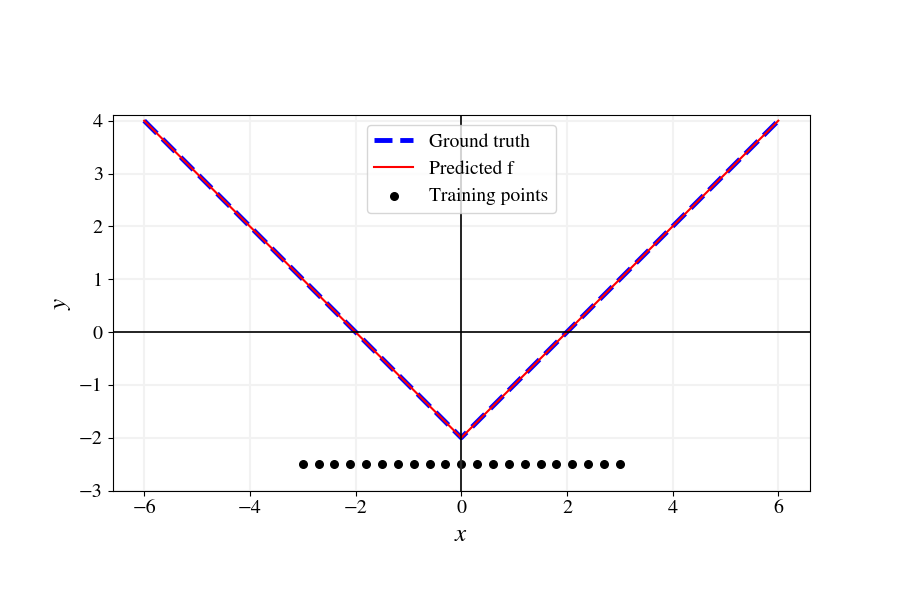}
    \caption{}\label{fig::groundTruthAbs}
\end{subfigure}%
\begin{subfigure}[b]{.5\linewidth}
\centering
    \centering
    \includegraphics[scale=0.35]{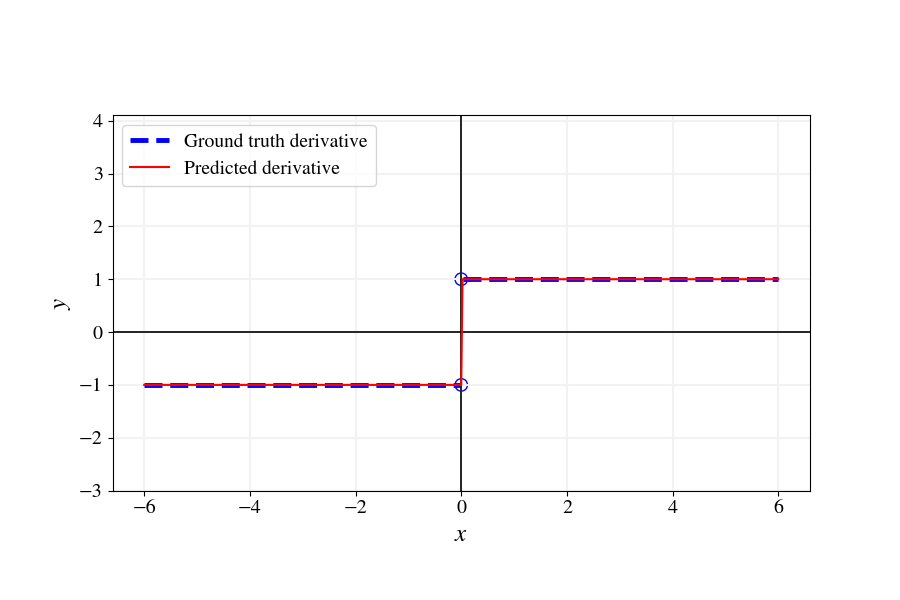}
    \caption{ }\label{fig::groundTruthAbsDeriv}
\end{subfigure}%
    \caption{Using $21$ equidistant training points between $-3$ and $3$  to train $f(x) = \abs{x} - 2$ with ICNN, (\protect\subref{fig::groundTruthAbs}) Ground truth, prediction and training point positions, (\protect\subref{fig::groundTruthAbsDeriv}) discontinuous ground derivative and approximated derivative with automatic differentiation.}
    \label{fig:}
\end{figure}

In this paragraph we quantitatively show that ReLU based input convex neural networks are able to very accurately train and predict data which includes vertices as well as allows for obtaining gradients at non-differentiable positions of these functions using subgradient techniques, see e.g. \cite{baydin2018automatic}. 
This allows us to accurately capture functions that include facets (such as yield functions) but still enables us to differentiate at these points.  
As an example we study the behavior of the trained ICNN
on data from the non-smooth but convex function
\begin{equation}
    f(x) = \abs{x} - 2
\end{equation}
where we use $21$ equidistant points in $[-3,3]$ to train a ICNN model. 
ReLU based neural networks are able to approximate any piecewise linear function with only one hidden layer
\citep{arora2016understanding}.
We therefore train this dataset with an ICNN using only $8$ trainable parameters employing an ADAM optimizer and a learning rate of $1e-4$ for $50,000$ iterations.
Figure \ref{fig::groundTruthAbs} plots the predictive fit of the neural network as well as the ground truth and highlights the positions of the $21$ training points. It can be seen that the neural network prediction agrees with the ground truth and that the neural network generalizes well. The network takes a value of $1.99999$ at the vertex position of $x=0$.
Hence, the neural network is not only accurate globally but also captures the corner value basically exactly.
However, we are interested in the proficiency of the network to approximate the derivative of $f(x)$. The ground truth is given by
\begin{equation}
    f'(x) = \begin{cases}
       -1, & \text{if } x<0 \\
       1, & \text{if } x>0.
    \end{cases}
\end{equation}
The function $f(x)$ is naturally non-differentiable at $x=0$.
Figure \ref{fig::groundTruthAbsDeriv} shows the ground truth $f'(x)$ as well as the the derivative of the neural network output with regards to the input using automatic differentiation. It can be seen that due to using subgradient techniques to approximate the derivatives, the neural network is able to determine a derivative of $x=0.0$. 

We highlight this with a simple example involving a Newton-Raphson loop using the vertex position ($x_{0} = 0.0$) as the starting point of the algorithm.
For this, assume that we would like to find one of the roots of $-f(x)$ which is approximated by $-\hat{f}(x)$. Utilizing the classical Newton-Raphson approach we can do this by using the updating formula
\begin{equation}
    x_{i+1} = x_{i} - \frac{-\hat{f}(x_{i})}{\hat{f}'(x_{i})}
\end{equation}
with $x_{0} = 0.0$. The results of this process over the first 3 iterations are shown in Table \ref{table:NRResults}. We can see that using the approximated derivative at the vertex, the Newton-Raphson loop is able to accurately converge towards the correct solution after the first iteration. \\
Hence, using ICNNs for yield function fitting we are (potentially) not only able to correctly approximate vertices present in convex functions but also able to use automatic differentiation to reliably determine derivatives at technically non-differentiable points. This allows us to directly incorporate the trained yield surfaces in classical time-integration loops. This is a significant advantages in comparison to classical phenomenological approaches which either have to smooth out the fit or integrate subdifferential formulations "by hand" into Newton-Raphson loops, see \cite{de2011computational} for an example.

\begin{table}[h!]
    \centering
\begin{tabular}{|l|c|c|}
\hline 
Iteration $i$ & $x_{i}$ & $\abs{\hat{f}(x_{i})}$ \\ 
\hline 
\hline 
0 & 0.00000 &1.99999 \\ 
\hline 
1 & -1.99999 &0.00000 \\ 
\hline 
2 & -1.99999 &0.00000 \\ 
\hline 
3 & -1.99999 &0.00000 \\ 
\hline 
\end{tabular} 
\caption{Convergence of Newton-Raphson loop from "discontinuous" starting point at $x_{0}=0$.}
\label{table:NRResults}
\end{table}

\newpage
%% New version of the num-names style
\bibliography{bib.bib}

\end{document}